\documentclass[aps,preprint,tightenlines,floatfix,superscriptaddress,nofootinbib]{revtex4}
\pdfoutput=1
\usepackage{graphicx}
\usepackage{dcolumn}
\usepackage{amsfonts}
\def\beq{\begin{equation}}
\def\eeq{\end{equation}}
\def\bea{\begin{eqnarray}}
\def\eea{\end{eqnarray}}
\def\nn{\nonumber}

\def\roughly#1{\mathrel{\raise.3ex\hbox
{$#1$\kern-.75em\lower1ex\hbox{$\sim$}}}}
\def\lsim{\roughly<}

\def\btos{b \to s}
\begin{document}
\bibliographystyle{apsrev}

\preprint{\vbox {\hbox{UdeM-GPP-TH-11-200}}}

\vspace*{2cm}

\title{\boldmath Using
$t\to b \overline{b} c$ to Search for New Physics}

\def\umontreal{\affiliation{\it Physique des Particules, Universit\'e
    de Montr\'eal, \\ C.P. 6128, succ. centre-ville, Montr\'eal, QC,
    Canada H3C 3J7}}
\def\tayloru{\affiliation{\it Physics and Engineering Department,
    Taylor University, \\ 236 West Reade Ave., Upland, IN 46989, USA}}
\def\laplata{\affiliation{\it IFLP, CONICET -- Dpto. de F\'{\i}sica,
    Universidad Nacional de La Plata, C.C. 67, 1900 La Plata,
    Argentina}}

\tayloru
\umontreal
\laplata

\author{Ken Kiers}
\email{knkiers@taylor.edu}
\tayloru

\author{Tal Knighton}
\email{tal_knighton@taylor.edu}
\tayloru

\author{David London}
\email{london@lps.umontreal.ca}
\umontreal

\author{Matthew Russell} 
\email{russell2@math.rutgers.edu}
\altaffiliation{Current address: Department of Mathematics -- Hill
  Center, Rutgers, The State University Of New Jersey, 110
  Frelinghuysen Rd.,  Piscataway, NJ 08854-8019, USA.}  
\tayloru

\author{Alejandro Szynkman}
\email{szynkman@fisica.unlp.edu.ar}
\umontreal
\laplata

\author{Kari Webster}
\email{kari_webster@taylor.edu}
\tayloru

\date{\today}

\begin{abstract}
We consider new-physics (NP) contributions to the decay $t\to
b\overline{b}c$. We parameterize the NP couplings by an effective
Lagrangian consisting of 10 Lorentz structures. We show that the
presence of NP can be detected through the measurement of the partial
width. A partial identification of the NP can be achieved through the
measurements of a forward-backward-like asymmetry, a
top-quark-spin-dependent asymmetry, the partial rate asymmetry, and a
triple-product asymmetry. These observables, which vanish in the
standard model, can all take values in the 10-20\% range in the
presence of NP.
Since $\left|V_{tb}V_{cb}\right| \simeq \left|V_{ts} V_{cs}\right|$, 
most of our results also hold, with small changes, for 
$t\to s \overline{s}c$.
\end{abstract}

%\pacs{}

\maketitle

%\newpage
%\setcounter{page}{1}

\section{Introduction}

On the whole, measurements of observables in the $B$ system agree with
the Standard Model (SM). However, some cracks have started to appear.
There are now several quantities whose measured values differ from the
predictions of the SM. Although these disagreements are not
statistically significant -- they are typically at the level of $\sim
2\sigma$ -- they are intriguing since there are a number of different
$B$ decays and effects involved, and they all appear in $\btos$
transitions. Because of this, there have been numerous papers
examining new-physics (NP) flavour-changing neutral-current (FCNC)
contributions to the various $\btos$ processes. These analyses have
been performed in the context of specific NP models, or
model-independently.

In general, such NP can also contribute to FCNC processes involving
the top quark. This has been looked at, though much less so than in
$B$ decays. However, given that the LHC will produce a large number of
top quarks and will be able to measure flavour-changing $t$ decays, it
is important to explore the possibility of NP contributions to FCNCs
in the top sector. In the past, analyses have focused on rare top
decays such as $t\to c V$ ($V=g,\gamma,Z$) and $t\to c
h$~\cite{tcVrefs, diazcruz}.  Other top decays where NP effects have
been examined include $t\to b\tau^+\nu$~\cite{diazcruz, btaunu1, cruz,
  liu, atwood1994, arens, bidai, sonireview} and $t\to W^+
d_k$~\cite{tWd}.

In this paper we examine the decay $t\to b\overline{b}c$. In the SM,
this decay occurs at tree level, via $t\to b W\to b \overline{b}c$. On
the other hand, because it involves the small element $V_{cb}$
($\simeq 0.04$) of the Cabibbo-Kobayashi-Maskawa (CKM) quark mixing
matrix, the amplitude for this process is also rather small, and is
therefore quite sensitive to NP.  For example, there could be NP
contributions to this decay in models with a charged Higgs
boson~\cite{diazcruz}, or via $t\to X^0 c\to b \overline{b} c$, where
$X^0$ corresponds to some neutral particle (such as a $Z^\prime$ or a
non-SM Higgs boson). Such processes could interfere with the SM
process, leading to observable consequences, even if the intermediate
NP particle were heavier than the top quark. Rather than restricting
our attention to any one particular model, we examine NP contributions
to $t\to b \overline{b} c$ model-independently (i.e.,\ using an
effective Lagrangian).

Our model-independent treatment of $t\to b\overline{b}c$ takes into
account the effects of the 10 possible four-Fermi operators. These
operators contribute to both CP-conserving and CP-violating
observables.  For the CP-even observables, we consider the CP-averaged
partial width, a forward-backward-like asymmetry, and an asymmetry
that depends on the spin of the top quark. For the CP-odd observables,
we note that the decay $t\to b \overline{b} c$ is dominated by one
amplitude in the SM; i.e., there is only one weak phase involved. As
such, all CP-violating asymmetries are very suppressed in the SM, so
the observation of a non-zero asymmetry would be a smoking-gun signal
of NP. In this paper, we discuss two types of CP-odd asymmetries:
the partial-rate asymmetry (PRA) and a triple-product asymmetry (TPA).

PRAs require a strong phase in order to be non-zero.  Strong phases
can arise due to gluon exchange, but it is expected that such phases
will be small since the energies involved are so large.  Another
source of a strong phase is the width of the $W$.  In our calculation,
we ignore QCD-based strong phases and assume that the required strong
phase is due entirely to the width of the $W$. This means that only
SM-NP interference can lead to a PRA. On the other hand, in contrast
with PRAs, TPAs do not require a strong phase in order to be
non-zero. Thus, NP-NP interference terms can give rise to TPAs. As we
will see, TPAs generated by SM-NP interference tend to be small, but
NP-NP TPAs can be large. These are particularly interesting.

We show that the measurement of the partial width by itself can reveal
the presence of NP. However, if the NP exists, we will want to know
its identity, i.e.,\ which of the 10 operators is responsible, 
and the partial width 
measurement does not give us this information. In order to do this, it
is necessary to measure the other quantities mentioned above. Since
the various observables depend differently on the operators, the
knowledge of their sizes will give us an idea of which NP operators are
present. This will allow us in turn to deduce which model(s) might be
responsible for the observed effects.

Although we confine our attention to $t\to b\overline{b}c$ in this
work, we note that most of our results are easily transferable to
$t\to s\overline{s}c$ by the replacement $\left(b,\overline{b}\right)
\to \left(s,\overline{s}\right)$ in Feynman diagrams and expressions.
Since $\left|V_{tb}V_{cb}\right| \simeq \left|V_{ts} V_{cs}\right|$,
the branching ratio for $t\to s\overline{s}c$ is similar to that for
$t\to b\overline{b}c$, and the two processes would apriori have
similar sensitivities to NP effects.  One difference between $t\to
b\overline{b}c$ and $t\to s\overline{s}c$ is that the ``CPT''
correction to the PRA would be significant for $t\to s\overline{s}c$,
whereas it is miniscule for $t\to b\overline{b}c$ (see the discussion
in Sec.~\ref{sec:PRA} and Appendix~\ref{sec:CPT}).

The remainder of this paper is organized as follows.  In
Sec.~\ref{sec:SMNP} we write down the SM contribution to $t\to
b\overline{b}c$, and also parameterize NP contributions to this decay
in terms of an effective Lagrangian containing ten terms.  In
Sec.~\ref{sec:CPeven} (CP-even observables) we compute the CP-averaged
partial width for the decay under consideration, as well as a
forward-backward-like asymmetry and an asymmetry based on the spin of
the top quark.  The latter two asymmetries are both constructed in
such a way that they are zero within the context of the SM.  We close
this section with a brief numerical study, noting that the CP-even
asymmetries could reasonably be of order 10's of percent.
Section~\ref{sec:CPodd} contains our analysis of two CP-odd
observables -- the partial rate asymmetry and the triple-product
asymmetry.  Section~\ref{sec:discussion} concludes with a discussion
of our results.  Appendices~\ref{sec:AppA}, \ref{sec:CPT} and
\ref{sec:AppB} contain some technical details.  In particular,
Appendix~\ref{sec:CPT} contains a discussion of results related to the
CPT theorem, namely which vertex-type corrections must be considered
in the calculation of PRAs in order not to violate CPT.

\section{Standard Model and New-Physics Contributions}
\label{sec:SMNP}

In this section we parameterize the NP contributions to $t\to b
\overline{b}c$ in terms of an effective Lagrangian.  We then write
down expressions for the SM and NP amplitudes.  These expressions are
used in following sections to determine various CP-even and CP-odd
observables.

\begin{figure}[t]
\begin{center}
\resizebox{4in}{!}{\includegraphics*{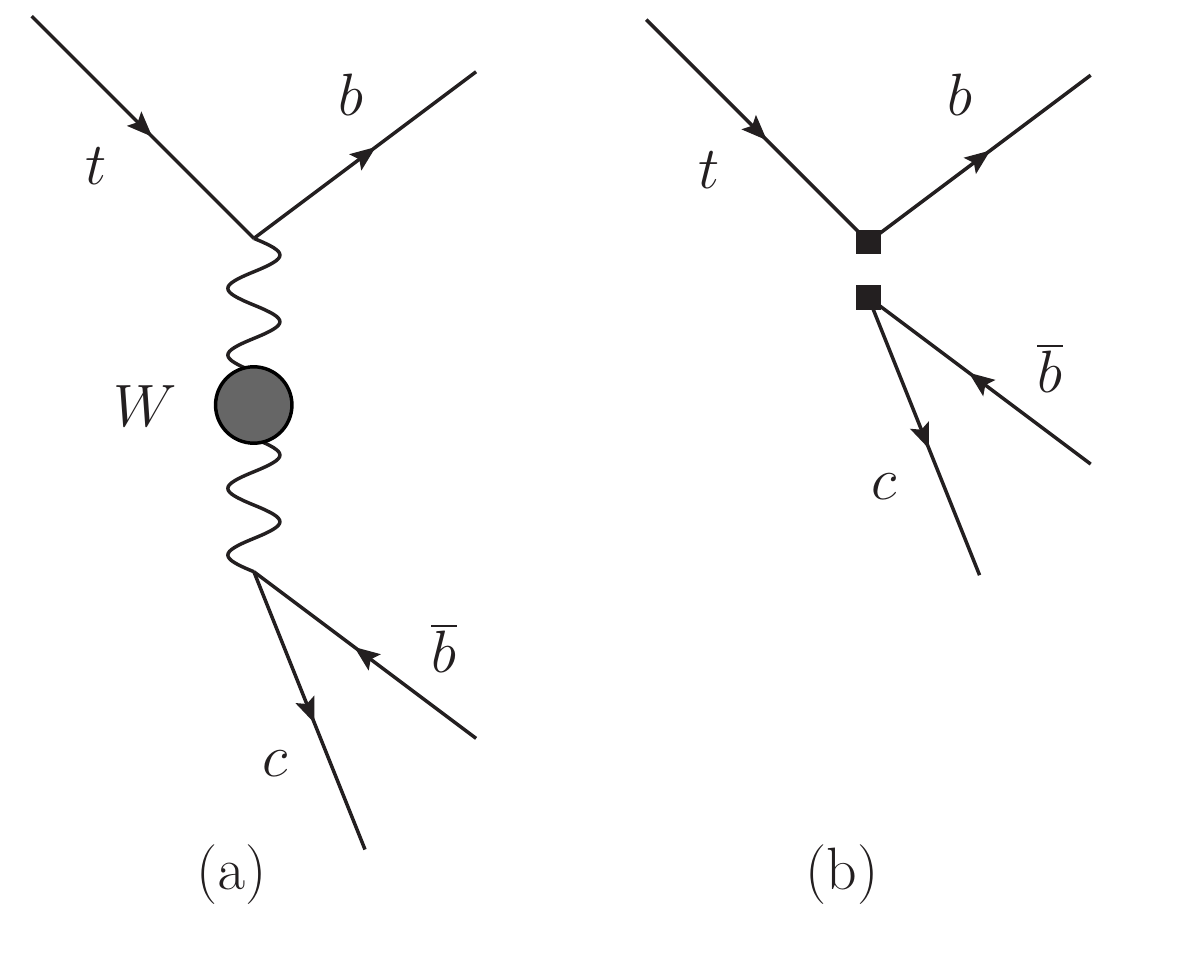}}
\caption{Feynman diagrams for $t\to b \overline{b}c$.  Diagram (a)
  shows the SM contribution.  Diagram (b) shows the NP contributions
  in the effective theory.  The NP contributions are assumed to have
  the same colour structure as that of the SM.  See
  Appendix~\ref{sec:AppB} for comments regarding the more general
  case.}
\label{fig:feynman_diagram1}
\end{center}
\end{figure}

Figure~\ref{fig:feynman_diagram1}(a) shows the Feynman diagram for the
SM contribution to $t\to b \overline{b} c$.  The resulting amplitude
is given by,
\begin{eqnarray}
  {\cal M}_W & = &
  -2\sqrt{2}G_F m_W^2 V_{cb}V_{tb}
    \left(\overline{u}_b\gamma_\alpha P_L u_t\right)\, 
    \left(\overline{u}_c \gamma_\beta P_L v_b\right) \left[ -g^{\alpha\beta} G_T(q^2) \right] ,~
   \label{eq:Wamp}
\end{eqnarray}
where $V$ is the CKM matrix. We work in the standard representation of
the CKM matrix, in which $V_{cb}$ and $V_{tb}$ are both real. Note that
colour indices have been suppressed.  The expression in square
parentheses is proportional to the $W$ propagator, with
$q=p_t-p_b=p_{\overline{b}}+p_c$,
$G_T(q^2)=\left[q^2-m_W^2+i\epsilon_T(q^2)\right]^{-1}$ and
$\epsilon_T(q^2) \simeq q^2 \Gamma_W/m_W$, where $\Gamma_W \simeq 3
G_F m_W^3/(2\sqrt{2}\pi)$.

(Note: throughout this paper we neglect the leptons' and light quarks'
masses. However, if this not done, the $W$ propagator is modified to
\beq
i\left[ \left( -g^{\alpha\beta} +\frac{q^\alpha q^\beta}{q^2} \right) G_T(q^2) + \frac{q^\alpha q^\beta}{q^2} G_L(q^2) \right] ~,
\eeq
where $G_L=\left[m_W^2+i\epsilon_L(q^2)\right]^{-1}$.
$\epsilon_T(q^2)$ and $\epsilon_L(q^2)$ are related to the transverse
and longitudinal widths of the $W$~\cite{atwood1994}, and they both
depend on the quark masses.  There has been considerable discussion in
the literature regarding the correct form of the $W$ propagator.
(See, for example, Refs.~\cite{atwood1994,nowakowski,liu,arens,castro,
  papavassiliou,PapavassiliouWeinbergModel,binosi_papavassiliou}.)
The above expression has been derived by performing a Dyson summation
of the absorptive parts of the $W$ self-energy diagrams in unitary
gauge, with quarks and leptons in the loops.  Some of the disagreement
in the literature has focused on the form of $G_L$ (see the brief
discussion in Ref.~\cite{sonireview}, for example). Still, when all
light masses are neglected, none of the observables in the present
work depend on $G_L$.  There seems to be broader agreement on the form
for $G_T$ in the literature, although many authors drop the $q^2$
dependence in $\epsilon_T$.  Finally, we should note that the Pinch
Technique may be used to reorganize perturbative calculations -- even
those involving resonances -- in such a way that results are
explicitly gauge-invariant (see, for example,
Ref.~\cite{binosi_papavassiliou}).  Rigorous application of the Pinch
Technique to the problem at hand is beyond the scope of this work.)

We parameterize new-physics effects via an effective Lagrangian ${\cal
  L}_{\mbox{\scriptsize eff}}={\cal L}_{\mbox{\scriptsize eff}}^V
+{\cal L}_{\mbox{\scriptsize eff}}^S+{\cal L}_{\mbox{\scriptsize
    eff}}^T$, where,
\begin{eqnarray}
  {\cal L}_{\mbox{\scriptsize eff}}^V & = & 
       \frac{g^{\prime 2}}{M^2}\left\{
    {\cal R}_{LL}^V\,\overline{b}\gamma_\mu P_L t \,
       \overline{c}\gamma^\mu P_L b
   + {\cal R}_{LR}^V\,\overline{b}\gamma_\mu P_L t \,
       \overline{c}\gamma^\mu P_R b
\right.\nonumber\\
& &~~~~\left.
   + {\cal R}_{RL}^V\,\overline{b}\gamma_\mu P_R t \,
       \overline{c}\gamma^\mu P_L b
   + {\cal R}_{RR}^V\,\overline{b}\gamma_\mu P_R t \,
       \overline{c}\gamma^\mu P_R b
\right\}+ \mbox{h.c.}, 
\label{eq:eff1}\\
&&\nonumber\\
  {\cal L}_{\mbox{\scriptsize eff}}^S & = & \frac{g^{\prime 2}}{M^2}\left\{
     {\cal R}_{LL}^S\,\overline{b} P_L t \,\overline{c} P_L b
   + {\cal R}_{LR}^S\,\overline{b} P_L t \,\overline{c} P_R b
\right. \nonumber\\
& &~~~~\left.
   + {\cal R}_{RL}^S\,\overline{b} P_R t \,\overline{c} P_L b
   + {\cal R}_{RR}^S\,\overline{b} P_R t \,\overline{c} P_R b
\right\}+\mbox{h.c.,} 
\label{eq:eff2}\\
&&\nonumber\\
  {\cal L}_{\mbox{\scriptsize eff}}^T & = & \frac{g^{\prime 2}}{M^2}\left\{
     {\cal C}_{T}\,\overline{b}\sigma_{\mu\nu}t \,
         \overline{c}\sigma^{\mu\nu}b
   + i \,{\cal C}_{TE}\,\overline{b}\sigma_{\mu\nu}t \,
         \overline{c}\sigma_{\alpha\beta}b
   \,\epsilon^{\mu\nu\alpha\beta}
\right\}+\mbox{h.c.}
\label{eq:eff3}
\end{eqnarray}
In the above expressions, $g^\prime$ is assumed to be of order $g$,
$M$ is the NP mass scale and the ${\cal R}$ and ${\cal C}$ couplings
may include weak (CP-violating) phases. For the Levi-Civita tensor, we
adopt the convention $\epsilon^{0123}=+1$.  The NP contributions to
$t\to b{\overline b}c$ are illustrated in
Fig.~\ref{fig:feynman_diagram1}(b).  Colour indices are not shown in
the above expressions, but are assumed to contract in the same manner
as those of the SM (i.e., $\overline{b}$ with $t$ and $\overline{c}$
with $b$).  In FCNC models -- those with a flavour-changing neutral
particle such as a $Z'$ or a scalar -- the colour indices would
contract in the opposite manner (i.e., $\overline{c}$ with $t$ and
$\overline{b}$ with $b$).  It is straightforward to incorporate
colour-mismatched terms into the effective Lagrangian.  This topic is
discussed further in Appendix~\ref{sec:AppB}.

It is useful to define
\begin{eqnarray}
  X^V_{LL} \equiv \left(\frac{g^\prime}{g}\right)^2
     \left(\frac{m_W}{M}\right)^2\frac{{\cal R}^V_{LL}}{V_{tb}V_{cb}}
     = \frac{\sqrt{2}}{8 G_F} \frac{g^{\prime 2}}{M^2}
       \frac{{\cal R}^V_{LL}}{V_{tb}V_{cb}}\, ,
\end{eqnarray}
and similarly for the other ${\cal R}$ and ${\cal C}$ couplings.  In
terms of the ``$X$'' parameters, we have the following expression for
the NP contribution to $t\to b\overline{b}c$,
\begin{eqnarray}
  {\cal M}_{\mbox{\scriptsize NP}} & = & 4\sqrt{2}G_F V_{cb}V_{tb}\left\{
    X_{LL}^V\,\overline{u}_b\gamma_\mu P_L u_t \,
        \overline{u}_c\gamma^\mu P_Lv_b
   + \ldots\right\}.
   \label{eq:NPamplitude2}
\end{eqnarray}
To get a sense of the possible order of magnitude of the $X$
couplings, note that, if $g^\prime \sim 2 g$ and $M\sim 500$~GeV, then
$X^V_{LL} \sim 2.5\times {\cal R}^V_{LL}$.  Thus, $X^V_{LL}$ could
reasonably be assumed to be of order unity. In other words, the SM and
NP contributions to $t\to b\overline{b}c$ can very well be about the
same size. When computing the effect of NP on a particular observable,
it is therefore important to include both the SM-NP interference and
NP$^2$ pieces.

\begin{figure}[t]
\begin{center}
\resizebox{2in}{!}{\includegraphics*{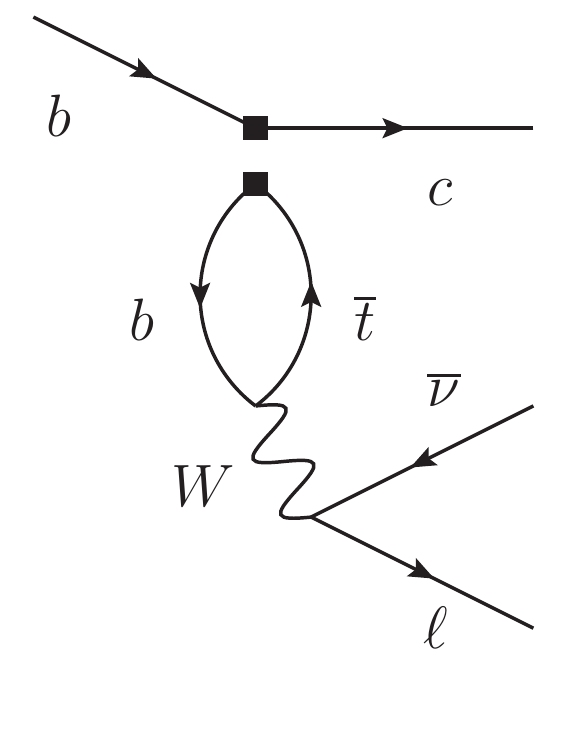}}
\caption{Loop-level contribution of the NP operators to $b\to c \ell
  \overline{\nu}$.  This contribution affects the measurement of
  $V_{cb}$.}
\label{fig:Vcb}
\end{center}
\end{figure}

At present, there are no direct constraints on the $X$ couplings.  The
precision measurements of $V_{cb}$ place an indirect constraint via
the loop diagram shown in Fig.~\ref{fig:Vcb}.  It is known that some
care must be taken when attempting to incorporate terms from an
effective Lagrangian into loop calculations \cite{BurgessLondon}.
Incorporating the diagram shown in Fig.~\ref{fig:Vcb}, we find the
following expression for the effective Lagrangian for $b\to c \ell
\overline{\nu}$,
\begin{eqnarray}
  {\cal L}_{\mbox{\scriptsize eff}}^{\mbox{\scriptsize SM+NP}} 
    \simeq -2\sqrt{2} \, G_F V_{cb}
    \left[
    \left(1+\zeta^V_{LL}\right)
     \overline{c}_L\gamma_\mu b_L
     + \zeta^V_{LR}\,\overline{c}_R\gamma_\mu b_R 
     \right]\overline{\ell}_L\gamma^\mu \nu_L + \mbox{h.c.,}
\label{eq:bcellnu}
\end{eqnarray}
in which we have dropped corrections of order ${\cal
  O}\left(m_b/m_t\right)$.  Using the Feynman rules for the various
vertices, and employing dimensional arguments, we estimate
\begin{eqnarray}
  \zeta^V_{LL(R)} &\sim & 
  \frac{G_F m_t^2}{2\sqrt{2}\pi^2}\left(V_{tb}\right)^2 
    X^V_{LL(R)}.
    \label{eq:zetaVLLR}
\end{eqnarray}
Since semileptonic $b\to c$ transitions are used to determine
$V_{cb}$, the experimental value of $V_{cb}$ can be used to bound
$X^V_{LL}$ and $X^V_{LR}$.  Let us first consider the $X^V_{LL}$ term
in Eq.~(\ref{eq:bcellnu}), ignoring the $X^V_{LR}$ term.  The
$X^V_{LL}$ term has exactly the same structure as the SM term.  Its
effect is thus simply to multiply any inclusive or exclusive $b\to c
\ell \overline{\nu}$ width by a factor of $\left[1 +
  2\mbox{Re}\left(\zeta^V_{LL}\right) +
  \left|\zeta^V_{LL}\right|^2\right]$.  The current experimental value
of $V_{cb}$ is $V_{cb} = \left(40.6\pm 1.3\right)\times
10^{-3}$~\cite{PDG}, implying a $6.4\%$ uncertainty on $V_{cb}^2$.  If
we assume that the $X^V_{LL}$ contribution to $b\to c \ell
\overline{\nu}$ is hiding in the experimental uncertainty of
$V_{cb}^2$, we find the bound,
\begin{eqnarray}
  \mbox{Re}\left(X^V_{LL}\right)\lsim 2.6 ,
\end{eqnarray}
in which we have neglected the quadratic contribution of
$\zeta^V_{LL}$, since it is small.  Since $X^V_{LR}$ is associated
with the right-handed quark current in Eq.~(\ref{eq:bcellnu}), its
effect is process-dependent.  For example, for $B\to D \ell
\overline{\nu}$, the hadronic matrix element is only sensitive to the
vector part of the hadronic current, so left-handed and right-handed
couplings both have the same effect, and they can be absorbed in with
the SM current~\cite{wu97}.  For other modes, such as $B\to D^*
\ell\overline{\nu}$, the right-handed and left-handed currents must be
treated differently~\cite{wu97}.  Since our expression in
Eq.~(\ref{eq:zetaVLLR}) is somewhat of an approximation in any case, we
assume the same upper bound for $X^V_{LR}$ as for $X^V_{LL}$, i.e.,
\begin{eqnarray}
  \mbox{Re}\left(X^V_{LR}\right)\lsim 2.6 .
\end{eqnarray}

The NP operators considered in this work could contribute, via loops,
to other observables as well.  As an example, consider the decay $B\to
\psi K_S$, which proceeds at tree-level in the SM.  In the present
context, the NP operators contribute to this decay via a diagram
similar to Fig.~\ref{fig:Vcb}, but with $\overline{\nu}\ell$ replaced
by $\overline{c}s$ in the final state.  The resulting effective
Lagrangian would be very similar to Eq.~(\ref{eq:bcellnu}), which
could lead to effects in the measurement of
$\sin(2\beta)$~\cite{kiers99}.  We do not consider such effects
further here.

Although we do not perform any model calculations in this work, it is
worthwhile to consider which types of models could give rise to the
various NP operators. The terms in Eqs.~(\ref{eq:eff1}) and (\ref{eq:eff2})
arise in models that contain new charged vector or scalar bosons. For
example, extensions of the SM containing gauge bosons with left- and
right-handed charged-current couplings (such as the Left-Right Model)
would contribute terms such as those appearing in ${\cal
  L}_{\mbox{\scriptsize eff}}^V$ -- including the ${\cal R}^V_{LR}$
and ${\cal R}^V_{RL}$ terms if there were some amount of mixing
between the left- and right-handed gauge bosons. Models containing
charged scalars (such as the charged Higgs bosons that appear in many
extensions of the SM) could give rise to the terms in the expression
for ${\cal L}_{\mbox{\scriptsize eff}}^S$. Alternatively, there are
many FCNC models containing a heavy neutral NP particle (such as a
$Z^\prime$ or a neutral Higgs boson) with flavour-changing $t$-$c$
couplings. Here Fierz rearrangements of the eight operator
combinations $\left(\gamma_\mu P_{L,R}\right)\left[\gamma^\mu
  P_{L,R}\right]$ and $\left(P_{L,R}\right)\left[P_{L,R}\right]$ (in
the notation employed in Ref.~\cite{nishi}) lead to all ten of the
operator combinations in Eqs.~(\ref{eq:eff1})-(\ref{eq:eff3}).  In
this case there would be mismatched colour indices between the NP and
SM diagrams.  Appendix~\ref{sec:AppB} explains how to deal with this
situation.

In the following sections we compute various observables associated
with the decays $t\to b \overline{b}c$ and $\overline{t}\to
\overline{b}b\overline{c}$.  We take as our starting point the
expressions for the SM and NP amplitudes in Eqs.~(\ref{eq:Wamp}) and
(\ref{eq:NPamplitude2}), respectively. 

\section{CP-even Observables}
\label{sec:CPeven}

We begin by considering three CP-even observables associated with the
decays in question.  The first is the CP-averaged partial width,
normalized to the SM result; the second is a forward-backward-like
asymmetry; and the third is a CP-even asymmetry that employs the spin
of the top quark.  In
Secs.~\ref{sec:CPaveragedwidth}-\ref{sec:SingleSpin} we work out
expressions for the various observables.
Section~\ref{sec:DiscussionCPeven} contains a numerical analysis and
discussion of the results.

\subsection{CP-averaged partial width}
\label{sec:CPaveragedwidth}

We first consider the partial width for $t\to b\overline{b}c$.
The expression for the differential partial width, including the
various NP terms from the effective Lagrangian, may be found in
Eq.~(\ref{eq:AppdGam}) in Appendix~\ref{sec:AppA}.  Performing the
integrations over
$\rho^2=\left(p_t-p_c\right)^2=\left(p_{\overline{b}}+p_b\right)^2$
and $q^2$, we find,
%n
\begin{eqnarray}
  && \Gamma\left(t\to b\overline{b} c\right)
   \simeq \Gamma_{\mbox{\scriptsize SM}}\left(t\to b\overline{b} c\right)
   \Bigg\{ 1 +\frac{4\Gamma_W}{m_W}\left[
     -0.04 \times\, \mbox{Re}\left(X_{LL}^{V*}\right) 
     + \mbox{Im}\left(X_{LL}^{V*}\right)\right] \nonumber\\
   && + ~\frac{3G_F m_t^2}{\sqrt{2}\pi^2
     \left(1-\zeta_W^2\right)^2\left(1+2\zeta_W^2\right)}
   \Bigg[
     \left|X_{LL}^V\right|^2 + \left|X_{LR}^V\right|^2 +
     \left|X_{RL}^V\right|^2 + \left|X_{RR}^V\right|^2 
     \nonumber\\
   && +~
     \frac{1}{4}\left(\left|X_{LL}^S\right|^2 + \left|X_{LR}^S\right|^2 +
     \left|X_{RL}^S\right|^2 + \left|X_{RR}^S\right|^2\right) +
     24\left|X_T\right|^2 + 96\left|X_{TE}\right|^2
     \Bigg]
   \Bigg\}\, ,
   \label{eq:Gam_tbbbarc}
\end{eqnarray}
where $\zeta_W \equiv m_W/m_t$ and
\begin{eqnarray}
  \Gamma_{\mbox{\scriptsize SM}}\left(t\to b\overline{b} c\right)
  \simeq \frac{G_F m_t^3}{24\sqrt{2}\pi}
   \left(V_{tb}V_{cb}\right)^2
   \left(1-\zeta_W^2\right)^2\left(1+2\zeta_W^2\right).
   \label{eq:GamSM}
\end{eqnarray}
In calculating the expressions in Eqs.~(\ref{eq:Gam_tbbbarc}) and
(\ref{eq:GamSM}), we have used the narrow width approximation, in
which $|G_T(q^2)|^2$ is replaced by a $\delta$-function in $q^2$,
appropriately normalized. [One exception is the term proportional to
  Re($X_{LL}^{V*}$) in Eq.~(\ref{eq:Gam_tbbbarc}), which was computed
  numerically.]

(Note that the term proportional to Im($X_{LL}^{V*}$) in
Eq.~(\ref{eq:Gam_tbbbarc}), which is involved in the partial rate
asymmetry discussed below, is not quite complete. In its present form,
it would lead to a violation of CPT.  To avoid running into problems
with the CPT theorem, certain vertex-type corrections need to be
included in the calculation when computing partial rate asymmetries.
We discuss these extra terms in Appendix \ref{sec:CPT}.)

The partial width for the CP-conjugate decay may be obtained from
Eq.~(\ref{eq:Gam_tbbbarc}) by complex conjugating all weak
phases.\footnote{Note that the ``$i$'' multiplying $X_{TE}$ in the NP
  amplitude does {\em not} get complex conjugated when computing the
  amplitude for the CP-conjugate process.}  This has the effect of
changing the sign of the Im($X_{LL}^{V*}$) term, while leaving the
other terms unchanged.  Adding the widths for $t\to b\overline{b}c$
and $\overline{t}\to \overline{b}b\overline{c}$, and dividing by twice
the SM result yields
\begin{eqnarray}
  {\cal R} & \equiv & \frac{\Gamma+\overline{\Gamma}}
  {2\Gamma_{\mbox{\scriptsize SM}}} \nonumber\\
  & \simeq & 1 + 0.0845\times 
   \Bigg[-0.05\!\times\! \mbox{Re}\left(X_{LL}^{V*}\right) +
     \left|X_{LL}^V\right|^2 + \left|X_{LR}^V\right|^2 +
     \left|X_{RL}^V\right|^2 + \left|X_{RR}^V\right|^2 
     \nonumber\\
   && +
     \frac{1}{4}\left(\left|X_{LL}^S\right|^2 + \left|X_{LR}^S\right|^2 +
     \left|X_{RL}^S\right|^2 + \left|X_{RR}^S\right|^2\right) +
     24\left|X_T\right|^2 + 96\left|X_{TE}\right|^2
     \Bigg] ,
   \label{eq:Rcal}
\end{eqnarray}
in which we have inserted the known values for the various physical
constants, and used the expression for $\Gamma_W$ noted below
Eq.~(\ref{eq:Wamp}).  Note that, in practice, the term proportional
to Re($X_{LL}^{V*}$) is always small compared to the other terms.

Above, we noted that the $X$'s could well be $O(1)$. Thus, from
Eq.~(\ref{eq:Rcal}), we see that the CP-averaged partial width could
be used as a tool to search for NP.  In particular, an experimental
value for ${\cal R}$ that is different from unity would give clear
evidence for NP. (Depending on the size of the NP signal, it may be
important to include corrections to Eq.~(\ref{eq:GamSM})
\cite{bernreuther}.)

On the other hand, all 10 NP operators contribute to ${\cal R}$ in
similar ways. Thus, even if the measurement of ${\cal R}$ reveals the
presence of NP, it does not give us any clue as to the type of NP. For
this reason, it is important to look for signs of NP in other
quantities. This is most easily done using observables that are
strictly zero within the context of the SM.  Such observables would
typically depend upon differing combinations of NP parameters, so that
the observation of NP effects using several different observables
would yield insight into the precise nature of the NP. In the
following, we construct several asymmetries that are zero within the
context of the SM and discuss their potential usefulness as tools for
searching for NP.

Note that the ratio ${\cal R}$, defined above, will also appear in the
denominators of all asymmetries considered below.  Since ${\cal R}$ is
primarily a sum of positive quantities, and since it will always
appear in the denominators of the asymmetries, it will always tend to
decrease the values of the asymmetries compared to the analogous
expressions employing the approximation ${\cal R}\approx 1$.

\subsection{Forward-Backward-Like Asymmetry}
\label{sec:FBlikeasym}

A tool that has historically been useful to experimentalists is the
forward-backward (FB) asymmetry.  The differential width for $t\to b
\overline{b}c$ may be written in terms of
$q^2=\left(p_t-p_b\right)^2=\left(p_{\overline{b}}+p_c\right)^2$ and
$\cos\theta$, where $\theta$ is the angle between the momentum of the
top quark and that of the charm quark in the $\overline{b}$-$c$ rest
frame.  The FB asymmetry makes use of the following asymmetric
integration over $\cos\theta$,
\begin{eqnarray}
  \Gamma_{\mbox{\scriptsize FB}} = \int_0^{m_t^2} \left[
    \int_0^1\frac{d\Gamma}{dq^2 \,d\!\cos\theta} d\!\cos\theta
    -\int_{-1}^0\frac{d\Gamma}{dq^2\, d\!\cos\theta} d\!\cos\theta\right]
  dq^2 \, .
  \label{eq:FB1}
\end{eqnarray}
We choose instead to work with the kinematical variables $q^2$ and
$\rho^2$, noting that
\begin{eqnarray}
  \cos\theta = \frac{m_t^2-2\rho^2-q^2}{q^2-m_t^2},
  \label{eq:costheta}
\end{eqnarray}
Using Eq.~(\ref{eq:costheta}), we may rewrite Eq.~(\ref{eq:FB1}) as
follows,
\begin{eqnarray}
  \Gamma_{\mbox{\scriptsize FB}} = \int_0^{m_t^2} \left[
    \int_{\left(m_t^2-q^2\right)/2}^{m_t^2-q^2}
      \frac{d\Gamma}{dq^2 d\rho^2} d\rho^2
    -\int_{0}^{\left(m_t^2-q^2\right)/2}
      \frac{d\Gamma}{dq^2 d\rho^2} d\rho^2\right]
  dq^2 \, .
  \label{eq:FBdef}
\end{eqnarray}
\begin{figure}[t]
\begin{center}
\resizebox{4in}{!}{\includegraphics*{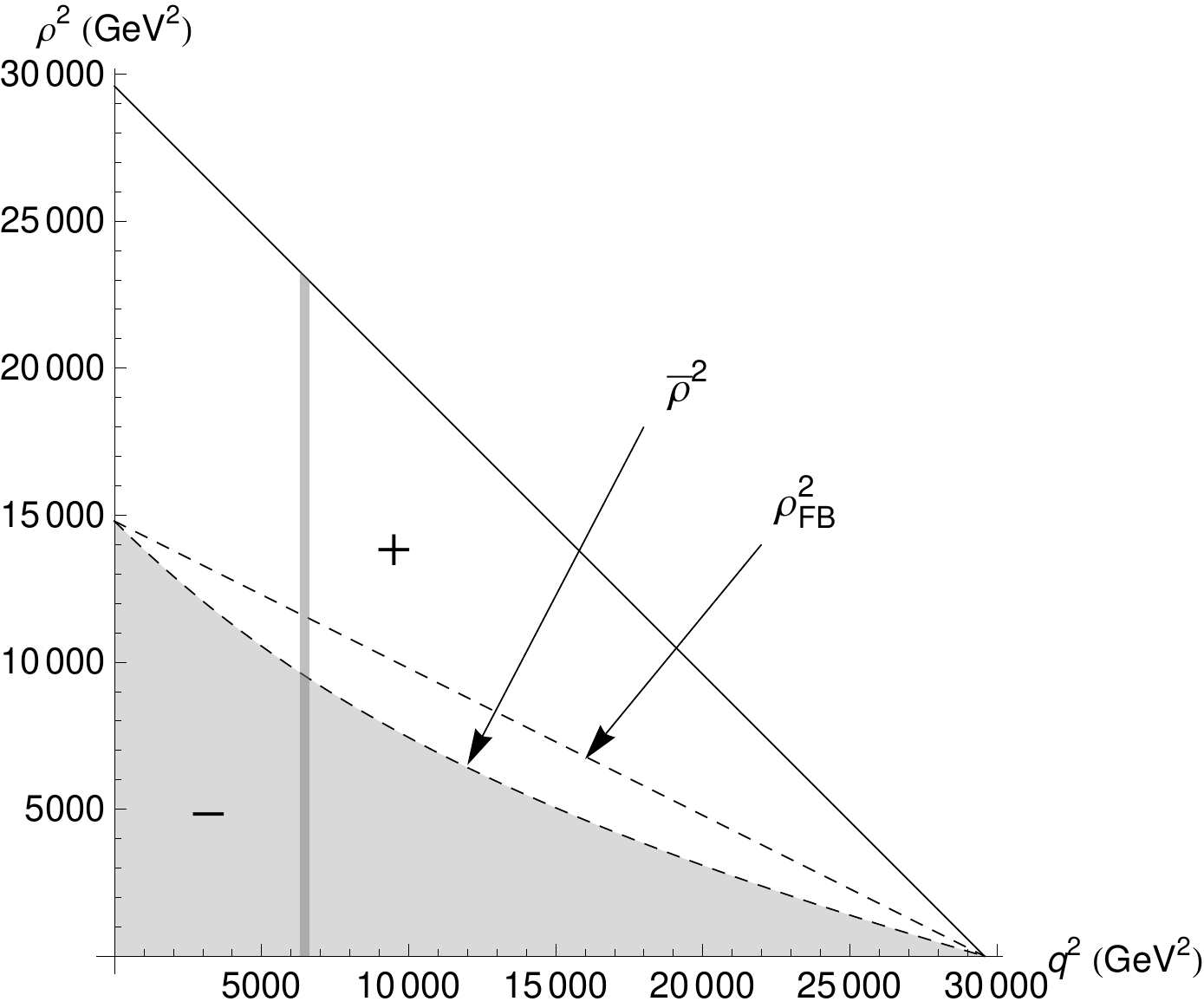}}
\caption{Phase space for $t\to b \overline{b} c$.  The gray vertical
  bar shows the location of the $W$ resonance at $q^2=m_W^2$.  The
  shaded and clear regions (separated by the curve denoted
  ``$\overline{\rho}^2$'') are used in the construction of the FB-like
  asymmetry $A_{\overline{\rho}^2}$.  The line indicated by
  ``$\rho^2_{\mbox{\scriptsize FB}}$'' shows the boundary used in the
  usual definition of the FB asymmetry.}
\label{fig:phasespace}
\end{center}
\end{figure}
Let us first consider the SM contribution to the FB asymmetry.  The
SM-only contribution to the width is such that
\begin{eqnarray}
  \frac{d\Gamma_{\mbox{\scriptsize SM}}}{dq^2 d\rho^2}
  \propto \left|G_{T}\left(q^2\right)\right|^2\left(q^2+\rho^2 \right)
    \left(m_t^2-q^2-\rho^2\right) 
    \label{eq:SMdqsqdrhosq}
\end{eqnarray}
(see Eq.~(\ref{eq:AppdGam}) in Appendix~\ref{sec:AppA}).  Using the
integration prescription in Eq.~(\ref{eq:FBdef}), and assuming the
narrow-width approximation, we find,
\begin{eqnarray}
  A_{\mbox{\scriptsize FB}}^{\mbox{\scriptsize SM}}
   = \frac{\Gamma_{\mbox{\scriptsize FB}}^{\mbox{\scriptsize SM}}+
    \overline{\Gamma}_{\mbox{\scriptsize FB}}^{\mbox{\scriptsize SM}}}
    {\Gamma_{\mbox{\scriptsize SM}}+\overline{\Gamma}_{\mbox{\scriptsize SM}}}
   = \frac{\Gamma_{\mbox{\scriptsize FB}}^{\mbox{\scriptsize SM}}}
    {\Gamma_{\mbox{\scriptsize SM}}} 
   \simeq -\frac{3\zeta_W^2}{2\left(1+2\zeta_W^2\right)}
   \simeq -0.228 .
   \label{eq:AFBSM}
\end{eqnarray}
Thus, we see that the SM contribution to the FB asymmetry is non-zero.

As noted above, in order to use a particular asymmetry as a
discriminator of NP, it is useful if the asymmetry is zero when no NP
contribution is present.  The regular FB asymmetry does not satisfy
this requirement, as is evidenced by Eq.~(\ref{eq:AFBSM}).  It turns
out, however, that if we modify the $\rho^2$ integration prescription
somewhat, the SM contribution can be made to disappear.  That is,
instead of breaking up the integral over $\rho^2$ at the point
$\rho^2_{\mbox{\scriptsize FB}}=\left(m_t^2-q^2\right)/2$, as is done
in Eq.~(\ref{eq:FBdef}), we move the boundary to the value
$\overline{\rho}^2$,
\begin{eqnarray}
  \Gamma_{\overline{\rho}^2} \equiv \int_0^{m_t^2} \left[
    \int_{\overline{\rho}^2}^{m_t^2-q^2}
      \frac{d\Gamma}{dq^2 d\rho^2} d\rho^2
    -\int_{0}^{\overline{\rho}^2}
      \frac{d\Gamma}{dq^2 d\rho^2} d\rho^2\right]
  dq^2 \, ,
  \label{eq:FB-likedGam}
\end{eqnarray}
in which $\overline{\rho}^2$ is chosen such that [see
Eq.~(\ref{eq:SMdqsqdrhosq})],
\begin{eqnarray}
    \int_{0}^{\overline{\rho}^2}
      \left(q^2+\rho^2 \right)\left(m_t^2-q^2-\rho^2\right) d\rho^2 
    =\int_{\overline{\rho}^2}^{m_t^2-q^2}
    \left(q^2+\rho^2 \right)\left(m_t^2-q^2-\rho^2\right) 
      d\rho^2.
  \label{eq:FB-likedGamSMzero}
\end{eqnarray}
Then, by construction,
\begin{eqnarray}
  \Gamma_{\overline{\rho}^2}^{\mbox{\scriptsize SM}} 
  = \int_0^{m_t^2} \left[
    \int_{\overline{\rho}^2}^{m_t^2-q^2}
      \frac{d\Gamma_{\mbox{\scriptsize SM}}}{dq^2 d\rho^2} d\rho^2
    -\int_{0}^{\overline{\rho}^2}
      \frac{d\Gamma_{\mbox{\scriptsize SM}}}{dq^2 d\rho^2} d\rho^2
      \right] dq^2 = 0 .
  \label{eq:FB-likedGamSMzero2}
\end{eqnarray}
The new boundary, $\overline{\rho}^2$, is $q^2$-dependent and can be
solved for numerically.\footnote{The equation for $\overline{\rho}^2$
  is cubic and can also be solved analytically, although the resulting
  expressions for the roots of the equation are not particularly
  enlightening.}  Figure~\ref{fig:phasespace} shows the phase space
available for $t\to b\overline{b}c$ and also indicates the two
boundary choices described above.  The vertical band indicates the
location of the $W$ resonance for the SM contribution.

Equation~(\ref{eq:AppdGam}) in Appendix~\ref{sec:AppA} gives the
expression for $d\Gamma/dq^2 d\rho^2$.  The $\rho^2$ dependence of the
SM piece is given by
$\left(q^2+\rho^2\right)\left(m_t^2-q^2-\rho^2\right)$; the SM-NP
cross terms and the $\left|X_{LL}^{V}\right|^2$ and
$\left|X_{RR}^{V}\right|^2$ terms have this same $\rho^2$ dependence.
Since the integration prescription described above is engineered to
eliminate the SM term when integrating over $\rho^2$, these latter
terms also disappear upon integration over $\rho^2$ in this manner.
Performing the integration numerically for the other terms yields the
following,
\begin{eqnarray}
  \Gamma_{\overline{\rho}^2} &\simeq &
  \frac{3G_F^2 m_t^5\left(V_{tb}V_{cb}\right)^2}{16\pi^3}
  \Bigg[ 
   0.155 \left(\left|X_{LR}^{V}\right|^2+\left|X_{RL}^{V}\right|^2\right)
   \nonumber\\
 && 
  +0.0208\left(\left|X_{LL}^{S}\right|^2+\left|X_{RR}^{S}\right|^2
      +\left|X_{LR}^{S}\right|^2+\left|X_{RL}^{S}\right|^2\right)
   \nonumber\\
 && 
  +0.310 \,\mbox{Re}\!\left[X_T\left(X_{LL}^{S*}+X_{RR}^{S*}\right)
     -2X_{TE}\left(X_{LL}^{S*}-X_{RR}^{S*}\right)\right] \nonumber\\
 &&+1.81
   \left(\left|X_T\right|^2+4\left|X_{TE}\right|^2\right)
   \Bigg] .
\label{eq:Gammarhobarsq}
\end{eqnarray}
The above expression is CP-even; i.e., the analogous expression for
$\overline{t}\to \overline{b}b\overline{c}$ is the same.
Finally, we form an FB-like asymmetry as follows,
\begin{eqnarray}
  A_{\overline{\rho}^2} = 
   \frac{\Gamma_{\overline{\rho}^2}+\overline{\Gamma}_{\overline{\rho}^2}}
    {\Gamma+\overline{\Gamma}}
  &\simeq & \frac{1}{\cal R}
\Bigg[ 
   0.0393 \left(\left|X_{LR}^{V}\right|^2+\left|X_{RL}^{V}\right|^2\right)
   \nonumber\\
 && 
  +0.00528\left(\left|X_{LL}^{S}\right|^2+\left|X_{RR}^{S}\right|^2
      +\left|X_{LR}^{S}\right|^2+\left|X_{RL}^{S}\right|^2\right)
   \nonumber\\
 && 
  +0.0786 \,\mbox{Re}\!\left[X_T\left(X_{LL}^{S*}+X_{RR}^{S*}\right)
     -2X_{TE}\left(X_{LL}^{S*}-X_{RR}^{S*}\right)\right] \nonumber\\
 &&+0.460
   \left(\left|X_T\right|^2+4\left|X_{TE}\right|^2\right)
   \Bigg] .
\label{eq:Arhobarsq}
\end{eqnarray}
By construction, this asymmetry is only non-zero if NP contributions
are present.  Section~\ref{sec:DiscussionCPeven} contains a discussion
of the range of sizes that are possible for the FB-like asymmetry.  At
this point we note only that asymmetries of order tens of percent are
possible.

\subsection{CP-even Spin Asymmetry}
\label{sec:SingleSpin}

The final CP-even observable that we consider depends on the spin of
the top quark~\cite{toppolarization}.  We construct this asymmetry in
such a way that it will be zero in the SM and potentially non-zero in
the context of NP.  Equation~(\ref{eq:absMsq_top_spin_only}) in
Appendix~\ref{sec:AppA} contains the expression for the absolute value
squared of the total amplitude, keeping only those terms that contain
the top-quark spin four-vector.  The term proportional to
$\left|G_T\right|^2$ in that expression is the SM contribution.  The
next term [proportional to Re$\left( G_T X^{V*}_{LL}\right)$] arises
from the interference of the SM contribution with one of the NP terms.
The remaining terms are purely NP in origin.  Inspection of
Eq.~(\ref{eq:absMsq_top_spin_only}) reveals that the SM term is
proportional to $p_{\overline{b}}\cdot s_t$.  (This is related to the
fact that, in the SM, the spin of the top is in the direction of the
momentum of the $\overline{b}$ in the top's rest
frame~\cite{soni1992}.)  Working in the top rest frame, we define
\begin{eqnarray}
  \vec{s}_{\parallel,\pm} = \pm \frac{1}{\sin\theta_{\overline{b}c}}
  \left(\widehat{n}_{c}
   -\widehat{n}_{\overline{b}}\cos\theta_{\overline{b}c}
  \right) \, ,
  \label{eq:sparallel}
\end{eqnarray}
where $\widehat{n}_{\overline{b}(c)}=\vec{p}_{\overline{b}(c)}/
|\vec{p}_{\overline{b}(c)}|$ and 
where $\theta_{\overline{b}c}$ is the angle (assumed to be between $0$
and $\pi$) between the three-momentum of the $\overline{b}$ and that
of the $c$, in the top's rest frame.  The cosine and sine of this
angle are given, respectively, by,
\begin{eqnarray}
  \cos\theta_{\overline{b}c} &=& \frac{m_t^2\left(\rho^2-q^2\right)
    -\rho^2\left(\rho^2+q^2\right)}
  {\left(m_t^2-\rho^2\right)\left(\rho^2+q^2\right)} ,\nn\\
  \sin\theta_{\overline{b}c} &=& \frac{2m_t\sqrt{\rho^2 q^2
      \left(m_t^2-q^2-\rho^2\right)}}
  {\left(m_t^2-\rho^2\right)\left(\rho^2+q^2\right)} .
  \label{eq:sintheta}
\end{eqnarray}
The vectors $\vec{s}_{\parallel,\pm}$ are in the decay plane and are
perpendicular to $\vec{p}_{\overline{b}}$ by construction.  Setting
$s_{t,\pm}^\mu = \left(0,\vec{s}_{\parallel,\pm}\right)$, we then have
$p_{\overline{b}}\cdot s_{t,\pm} =0$.  Thus, the SM contribution to
the amplitude squared disappears for these orientations of the top
quark's spin.  We can thus construct an asymmetry based on this spin
configuration that will be zero within the SM, making it potentially a
sensitive probe for NP.  We first define,
\begin{eqnarray}
  \Gamma_\parallel \equiv \frac{1}{2}\left[
    \Gamma\left(\vec{s}_{\parallel,+}\right)
     -\Gamma\left(\vec{s}_{\parallel,-}\right)\right],
\end{eqnarray}
where the factor of ``1/2'' is to account for the average over the
top quark's spins.  Using Eqs.~(\ref{eq:absMsq_top_spin_only}),
(\ref{eq:sparallel}) and (\ref{eq:sintheta}), and
incorporating the integration over phase space, we obtain,
\begin{eqnarray}
  \Gamma_\parallel &=&
  \frac{G_F^2 m_t^5\left(V_{tb}V_{cb}\right)^2}{70 \pi^2}
  \Bigg\{
  \left(\left|X_{LR}^{V}\right|^2-
       \left|X_{RL}^{V}\right|^2\right) \nonumber\\
 & & -\frac{1}{4}\left(\left|X_{LL}^{S}\right|^2-\left|X_{RR}^{S}\right|^2
      +\left|X_{LR}^{S}\right|^2-\left|X_{RL}^{S}\right|^2\right) \nonumber\\
 & &  +2\,\mbox{Re}\!\left[X_T\left(X^{S*}_{LL}-X^{S*}_{RR}\right)
       -2X_{TE}\left(X^{S*}_{LL}+X^{S*}_{RR}\right)\right]
  -96\,\mbox{Re}\!\left[X_T X_{TE}^*\right]\Bigg\} \, .
\end{eqnarray}
Finally, summing over the process and the CP-conjugate process, we
obtain,
\begin{eqnarray}
  A_\parallel &=& \frac{\Gamma_\parallel+\overline{\Gamma}_\parallel}
  {\Gamma+\overline{\Gamma}} \nonumber \\
  & \simeq & \frac{0.0607}{\cal{R}}
  \Bigg\{
  \left(\left|X_{LR}^{V}\right|^2-
       \left|X_{RL}^{V}\right|^2\right) 
 -\frac{1}{4}\left(\left|X_{LL}^{S}\right|^2-\left|X_{RR}^{S}\right|^2
      +\left|X_{LR}^{S}\right|^2-\left|X_{RL}^{S}\right|^2\right) \nonumber\\
 & &  +2\,\mbox{Re}\!\left[X_T\left(X^{S*}_{LL}-X^{S*}_{RR}\right)
       -2X_{TE}\left(X^{S*}_{LL}+X^{S*}_{RR}\right)\right]
  -96\,\mbox{Re}\!\left[X_T X_{TE}^*\right]\Bigg\} \, ,~~~
  \label{eq:Aparallel}
\end{eqnarray}
where we have used the fact that $12\sqrt{2}G_F m_t^2/[35 \pi
  (1-\zeta_W^2)^2(1+2\zeta_W^2)] \simeq 0.0607$.  A discussion of
numerical values obtainable for the CP-even single-spin asymmetry
follows in the next subsection.

\subsection{Discussion of CP-even Observables}
\label{sec:DiscussionCPeven}

In this section we have described three observables that are even
under CP and that could be used to detect the presence of NP in the
decays $t\to b \overline{b} c$ and
$\overline{t}\to\overline{b}b\overline{c}$.  Should NP be discovered,
detailed analysis of such observables could allow experimentalists to
map out the nature of the NP.

The first observable considered in this section was a ratio, ${\cal
  R}$, which was defined to be proportional to the CP-averaged partial
width.  Of the observables considered in this work, ${\cal R}$ would
probably be the simplest to measure experimentally.  Decisive
experimental deviation from the SM value ${\cal R}=1$ would be
evidence for NP. 

On the other hand, ${\cal R}$ cannot be used to distinguish the
different NP operators. To do this requires the use of other
quantities. It is useful to employ observables that give a null signal
within the context of the SM.  For such observables, a significant
departure from zero would be a ``smoking-gun'' signal for new physics.
In addition, since they depend differently on the various NP
operators, the observation of non-zero values for these observables
would help in identifying the type of NP.

The usual forward-backward asymmetry for $t\to b\overline{b}c$ is
expected to be non-zero within the context of the SM.  It is possible,
however, to alter the kinematical weighting that is used in defining
the FB asymmetry in such a way that the resulting ``FB-like''
asymmetry is zero within the context of the SM.
Equation~(\ref{eq:Arhobarsq}) defines the FB-like asymmetry
$A_{\overline{\rho}^2}$ in terms of an asymmetric integration over the
kinematical variable $\rho^2$.  The integration is engineered in such
a way that the SM contribution disappears kinematically.  In
Eq.~(\ref{eq:Aparallel}) we formed a CP-even asymmetry using the spin
of the top quark.  This asymmetry was also defined in such a way that
it was zero within the context of the SM.  A non-zero experimental
signal for either of these asymmetries would indicate the presence of
NP.

\begin{table}[t]
\caption{Some representative values for the FB-like asymmetry
  $A_{\overline{\rho}^2}$ and the CP-even spin asymmetry
  $A_\parallel$.  The value for ${\cal R}$ is also included for each
  case.}
\begin{tabular}{|cc|cccc|cc||c|c|c|}
\hline\hline
$X^V_{LR}$ & $X^V_{RL}$ & $X^S_{LL}$ & $X^S_{LR}$ &
$X^S_{RL}$ & $X^S_{RR}$ & $X_{T}$ & $X_{TE}$ & ${\cal R}$ 
& $A_{\overline{\rho}^2}$ & $A_{\parallel}$ \\
\hline
 ~$1.5$~ & 
   & & & & 
   & & 
    & ~$1.2$~ & ~$7.4\%$~ & ~$11\%$~ \\
  & 
   & & & $2.5$ & $2.5$ 
   & & 
    & $1.3$ & ~$5.2\%$~ & $15\%$ \\
  & 
   & ~$1$~ & & & ~$-1$~ 
   & ~$0.5$~ & ~$0.125$~
    & $1.7$ & ~$6.9\%$~ & $-14\%$ \\
  & 
   & ~$1$~ & & & ~$-1$~ 
   & ~$0.5$~ & ~$-0.125$~
    & $1.7$ & ~$12\%$~ & $29\%$ \\
  & 
   & $-2.5$ & & & $2.5$ 
   & & $0.25$
    & $1.8$  & ~$21\%$~ & $0\%$ \\
  & 
   & & & & 
   & ~$0.5$~& ~$-0.25$~
    & $2.0$ & ~$11\%$~ & $36\%$ \\
 ~$2.5$~ & $2.5$ 
   & & & &
   & & 
    & $2.1$ & ~$24\%$~ & $0\%$ \\
  & 
   & & & & 
   & $1$ & 
    & $3.0$ & ~$15\%$~ & $0\%$ \\
  & 
   & ~$2.5$~ & & & $2.5$
   & $1$ &
    & $3.3$ & ~$28\%$~ & $0\%$ \\
\hline\hline
\end{tabular}
\label{tab:ACPeven}
\end{table}

Table~\ref{tab:ACPeven} contains some representative values for the
CP-even asymmetries $A_{\overline{\rho}^2}$ and $A_\parallel$, along
with the corresponding value for the ratio ${\cal R}$ in each case.
The entries in the table are ordered from smaller ${\cal R}$ values in
the top rows to larger ones in the bottom rows.  As is evident from
the table, when ${\cal R}$ is close to unity (meaning that it may not
be a very clear discriminator of NP), it is still possible to have
CP-even asymmetries that are on the order of several percent.  For
larger ${\cal R}$ values, the asymmetries $A_{\overline{\rho}^2}$ and
$A_\parallel$ can reach into the 10's of percent.  Note, however, that
there are some NP scenarios in which $A_\parallel$ suffers
cancellations or is zero, even if the NP parameters themselves are
non-zero.  For example, if
$\left|X^V_{LR}\right|=\left|X^V_{RL}\right|$, the contributions from
these two parameters cancel in $A_\parallel$.

\begin{figure}[t]
\begin{center}
\resizebox{5.8in}{!}{\includegraphics*{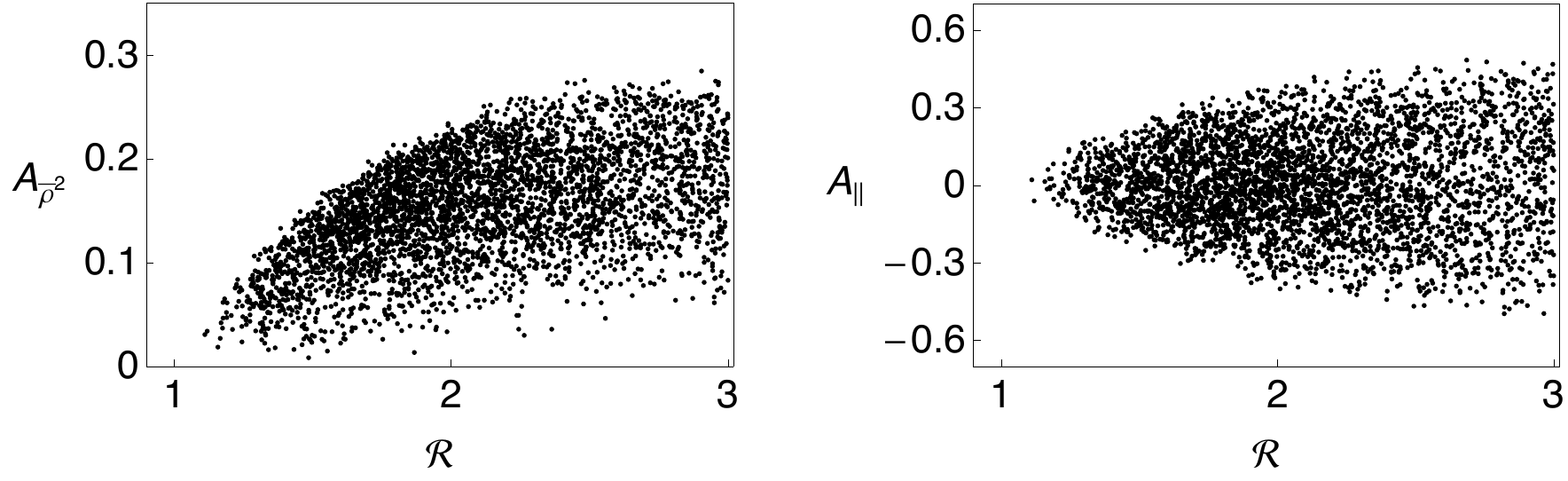}}
\caption{Scatter plots of the CP-even asymmetries
  $A_{\overline{\rho}^2}$ and $A_\parallel$ for various combinations
  of the NP parameters.}
\label{fig:ACPeven}
\end{center}
\end{figure}

The CP-even observables are displayed in another manner in
Fig.~\ref{fig:ACPeven}, which shows scatter plots of
$A_{\overline{\rho}^2}$ versus ${\cal R}$ and $A_\parallel$ versus
${\cal R}$.  The points in this plot were obtained by generating real
random numbers for eight of the ten NP parameters over various ranges.
($X^V_{LL}$ and $X^V_{RR}$ were excluded, since they do not contribute
to the numerator of either asymmetry.)  Asymmetries were only plotted
if ${\cal R}\leq 3$.  Again, it is evident that CP-even asymmetries of
order a few 10's of percent are possible.

\section{CP-odd Observables}
\label{sec:CPodd}

In addition to the CP-even observables considered in the previous
section, it is also possible to construct CP-odd observables related
to the decay $t\to b \overline{b} c$.  In this section we consider two
such observables.  The first is the partial rate asymmetry, which
compares the partial width for the process to that of the CP-conjugate
process.  The second is a triple-product asymmetry, which is formed
using the spin of the top quark and the three-momenta of two of the
final-state quarks.  To be non-zero, both of these asymmetries require
the presence of at least two contributing amplitudes with a
non-trivial relative weak phase.  Let us first consider the partial
rate asymmetry.

\subsection{Partial Rate Asymmetry}
\label{sec:PRA}

The SM amplitude for $t\to b \overline{b} c$ is dominated by a single
contribution. As such, the partial rate asymmetry vanishes.  In the
presence of NP, the partial-rate asymmetry (PRA) can be nonzero if
there is a NP contribution to the decay with a relative weak phase. As
can be seen in Eq.~(\ref{eq:Gam_tbbbarc}), there is one important NP
piece of this type -- $X^V_{LL}$.  The contribution to the PRA then
comes from the interference of the SM $W$-exchange amplitude with the
$X^V_{LL}$ term.  What we see in this subsection is that the PRA can
actually be of order several percent if the Lorentz structure of the
NP is $(V-A)\times(V-A)$.

We have noted above that a non-zero PRA requires the interference of
at least two amplitudes having a non-zero relative weak phase.
Another requirement is that these amplitudes have a non-zero relative
strong phase.  Strong phases can come from the exchange of gluons, but
they can also emerge from the imaginary parts of loop diagrams that do
not involve gluons.  In particular, if an exchanged particle in the
process has a resonance, there is a strong phase associated with the
width of that particle.  Strong phases originating from particles'
widths have been used to generate PRAs in many different systems,
including $t\to b\overline{b}c$~\cite{Eilam,diazcruz}, $t\to
b\tau^+\nu$~\cite{cruz,liu,diazcruz,atwood1994,arens,sonireview}, and
various supersymmetric decays~\cite{SUSY1,SUSY2,SUSY3}.  In the
present calculation, the width of the $W$ provides the required strong
phase.  This means that the PRA can only arise from SM-NP
interference, since NP-NP interference terms do not have a relative
strong phase.

Using the expression in Eq.~(\ref{eq:Gam_tbbbarc}), and recalling that
the analogous expression for the CP-conjugate process involves the
complex conjugation of the weak phases, we immediately find the
following expression for the partial rate asymmetry,
\begin{eqnarray}
  A_{\mbox{\scriptsize CP}} = 
   \frac{\Gamma-\overline{\Gamma}}{\Gamma+\overline{\Gamma}}
   \simeq \frac{1}{\cal R} \frac{4\Gamma_W}{m_W}
   \mbox{Im}\left(X_{LL}^{V*}\right) \simeq \frac{0.102}{\cal R}\times
   \mbox{Im}\left(X_{LL}^{V*}\right) .
   \label{eq:PRA1}
\end{eqnarray}
As was noted above, the PRA requires the existence of a non-zero
relative strong phase between interfering amplitudes.  In this
example, the strong phase is provided by the width of the $W$, which
is the reason that the PRA is proportional to $\Gamma_W$.  Examination
of Eq.~(\ref{eq:PRA1}) reveals that the best-case scenario for the PRA
occurs when $X_{LL}^V$ is purely imaginary, or nearly so, and all other
NP coefficients are zero.  In this case, ${\cal R} \simeq 1+0.0845
\left|X_{LL}^V\right|^2$, and we find that the PRA is maximized when
$\left|X_{LL}^V\right|\simeq 3.44$.  The maximum possible PRA,
obtained in this manner, is approximately 18\%.

\begin{figure}[t]
\begin{center}
\resizebox{1.7in}{!}{\includegraphics*{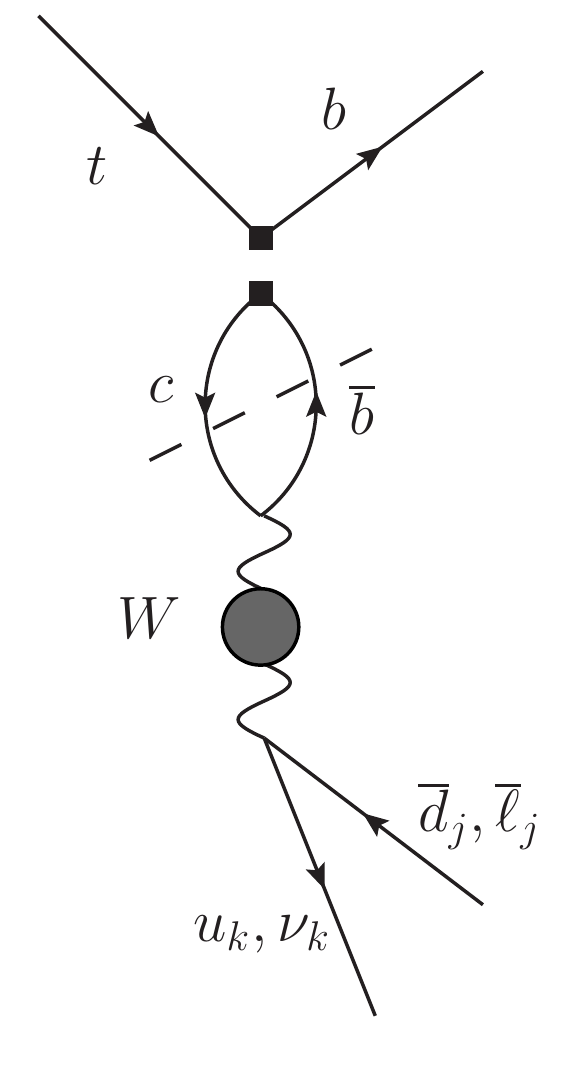}}
\caption{Vertex correction-type diagrams involving the effective
  operators shown in Fig.~\ref{fig:feynman_diagram1} (b).  These
  diagrams contribute to the cancellations required by the CPT
  theorem. The dashed line indicates that only the absorptive parts of
  the diagrams are computed.}
\label{fig:CPT_vertex_correction}
\end{center}
\end{figure}

As is well-known~\cite{gerard1988, wolfenstein1990}, the CPT theorem
requires that we actually be a bit more careful when computing PRAs.
In particular, invariance under CPT requires that the total width of
the top be equal to that of the anti-top.  Our result in
Eq.~(\ref{eq:PRA1}) shows that, under certain circumstances, the {\em
  partial} width for $t\to b \overline{b}c$ is not equal to the
partial width for $\overline{t}\to \overline{b}b\overline{c}$.  This
necessarily implies that there must be compensating partial rate
asymmetries in other top/anti-top decay modes such that the total top
width is still equal to the total anti-top width.  In order to respect
CPT in this way, it turns out that we need to include another class of
diagrams, shown in Fig.~\ref{fig:CPT_vertex_correction}.  These
diagrams contribute to various top decay modes, inducing partial rate
asymmetries in these modes in such a way that the total top width is
equal to the total anti-top width.  In the special case $t\to b
\overline{b}c$, the effect is such that ``$\Gamma_W$'' in the
numerator of Eq.~(\ref{eq:PRA1}) gets replaced by
``$\Gamma_W-\Gamma\left(W\to \overline{b}c\right)$''~\cite{Eilam,
  soares, Eilamreply}, which is to say that the strong phase due to
the rescattering process $W\to \overline{b}c\to W$ does not contribute
to the PRA.  Since $\Gamma\left(W\to \overline{b}c\right)$ is very
small, we may safely neglect its effect.  It is worthwhile to explore
this point a bit further, however, and we do so in Appendix~\ref{sec:CPT}.
Specifically, we verify that the diagrams in
Fig.~\ref{fig:CPT_vertex_correction} interfere with their SM
counterparts in such a way that the CPT theorem is respected, and we
also comment on the PRAs that result in other decay modes due to the
NP effective operators for $t\to b\overline{b}c$.

\subsection{Triple Product Asymmetry}
\label{sec:TPAsym}

Mathematically, triple-product asymmetries (TPAs) in $t\to b
\overline{b} c$ are related to terms of the form
$\vec{v}_i\cdot\left(\vec{v}_j\times\vec{v}_k\right)$ in the absolute
value squared of the amplitude, where each of the $\vec{v}_i$ could
represent a momentum or spin.  Working in the rest frame of the top
quark, there are only two independent three-momenta. Thus, in order to
obtain a non-zero TPA, we need to include one or more spins in
$\vec{v}_i\cdot\left(\vec{v}_j\times\vec{v}_k\right)$.  Since the
light final-state quarks hadronize, it is difficult to gain useful
information from their spins.  The situation is different for the top
quark, however, since it decays too quickly to hadronize.  In this
case, we can construct asymmetries based on
$\vec{s}\cdot\left(\vec{p}_1\times\vec{p}_2\right)$, where $\vec{s}$
is the top quark's spin and $\vec{p}_1$ and $\vec{p}_2$ are two of the
final-state momenta~\cite{kane}.  In the context of the calculation,
these terms arise from expressions such as
$\epsilon_{\alpha\beta\gamma\delta} p_t^\alpha s^\beta_t
p_{\overline{b}}^\gamma p_c^\delta$.

Now, the PRA considered above contained a factor of $\Gamma_W\sim
2$~GeV in the numerator.  The presence of this factor was due to the
requirement that there be a relative strong phase between diagrams
contributing to the PRA. On the other hand, TPAs do not require a
strong phase and are thus not suppressed by a factor of $\Gamma_W$.
This means that TPAs could in principle be much larger than the PRA
considered above. As we shall see, there are in fact certain NP
operators that can produce a large TPA.

Because TPAs are CP-odd quantities, they require a non-zero relative
weak phase between interfering diagrams, just as the PRA did.  But
because no strong phase is necessary, TPAs can in principle arise both
from SM-NP and NP-NP interference. (Due to the strong phase
requirement, the PRA could only arise from SM-NP interference.) What we
find, however, is that the only TPA that survives is one due to NP-NP
interference.

All triple-product terms in the absolute value squared of the
amplitude may be written in terms of
$\epsilon_{\alpha\beta\gamma\delta} p_t^\alpha s^\beta_t
p_{\overline{b}}^\gamma p_c^\delta$.  Keeping only such terms, we find
the following expression in the rest frame of the top quark,
\begin{eqnarray}
 &&\frac{1}{3}  
   \sum_{\mbox{\scriptsize colours}}
   \sum_{~b, \overline{b}, c\mbox{\scriptsize ~spins}}
   \left. \left|{\cal M}\right|^2
     \right|_{\mbox{\scriptsize TP}} \nonumber\\
     &&= 1536 G_F^2m_t^2\left(V_{tb}V_{cb}\right)^2 
     \vec{s}\cdot\left(\vec{p}_{\overline{b}}\times \vec{p}_c\right)
     \mbox{Im}\!\left[X_T\!\left(X^{S*}_{LL}\!+\!X^{S*}_{RR}\right)\!
       -\! 2X_{TE}\!\left(X^{S*}_{LL}\!-\!X^{S*}_{RR}\right)\right]\!, ~~~~~
      \label{eq:MstarMNPNP}
\end{eqnarray}
in which $\vec{s}$ denotes the top's spin [see also
Eq.~(\ref{eq:absMsq_top_spin_only})].  In computing the above
expression, we have summed over quark colours and over the final-state
quarks' spins, and have divided by 3 for the average over the top
quark's colours.  Setting
\begin{eqnarray}
  \vec{s}_{\perp,\pm} = \pm \frac{\vec{p}_{\overline{b}}\times \vec{p}_c}
      {\left|\vec{p}_{\overline{b}}\times \vec{p}_c\right|},
\end{eqnarray}
we define
\begin{eqnarray}
  \Gamma_{\mbox{\scriptsize TP}} \equiv 
  \frac{1}{2}\left[\Gamma\left(\vec{s}_{\perp,+}\right)
    - \Gamma\left(\vec{s}_{\perp,-}\right)\right] ,
  \label{eq:GammaTPdef}
\end{eqnarray}
where the factor of ``$1/2$'' is to account for the average over the
top quark's spins.  Using the result in Eq.~(\ref{eq:MstarMNPNP}) and
incorporating the integration over phase space, we obtain,
\begin{eqnarray}
  \Gamma_{\mbox{\scriptsize TP}}=
  \frac{2G_F^2 m_t^5\left(V_{tb}V_{cb}\right)^2}{35 \pi^2}
  \mbox{Im}\!\left[X_T\left(X^{S*}_{LL}+X^{S*}_{RR}\right)
       -2X_{TE}\left(X^{S*}_{LL}-X^{S*}_{RR}\right)\right]\!,~~~~~
          \label{eq:gamTPnon-suppressed}
\end{eqnarray}
in which we have used the fact that
\begin{eqnarray}
  \left|\vec{p}_{\overline{b}}\times \vec{p}_c\right|=
    \frac{1}{2 m_t}\left[q^2\rho^2\left(m_t^2-q^2-\rho^2\right)\right]^{1/2} ~.
\end{eqnarray}

\begin{table}[t]
\caption{Some representative values for the triple-product asymmetry.
  The second to last column also shows ${\cal R}$ for each case.}
\begin{tabular}{|cccc||c|c|}
\hline\hline
$X^S_{LL}$ & 
$X^S_{RR}$ & $X_{T}$ & $X_{TE}$ & ${\cal R}$ &
$A_{\mbox{\scriptsize CP}}^{\mbox{\scriptsize TP}}$\\
\hline
 $-1.5i$ & $-1.5i$ & 0.5 & 0 & 1.6 & 23\% \\
 $2.5i$ & $-2.5i$ & 0 & 0.4 & 2.6 & 38\% \\
 ~~~~~$-2.5i$~~~~~ & ~~~~~$-2.5i$~~~~~ & ~~~~~1~~~~~ & ~~~~~0~~~~~  
& ~~~~~3.3~~~~~ & ~~~~~37\%~~~~~ \\
\hline\hline
\end{tabular}
\label{tab:ACPTPNPNP}
\end{table}

\begin{figure}[t]
\begin{center}
\resizebox{3.1in}{!}{\includegraphics*{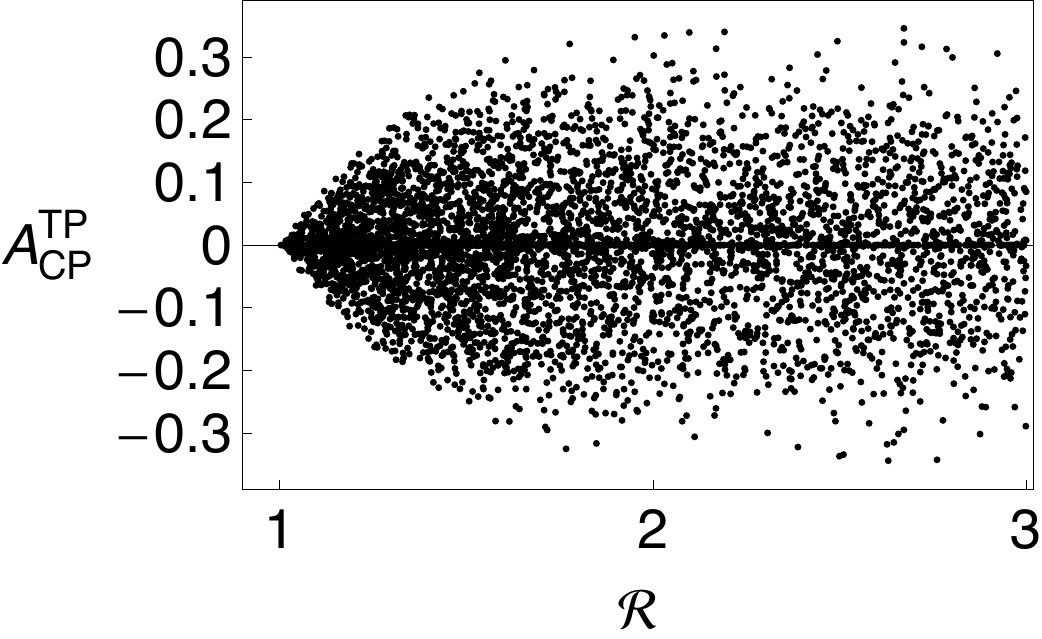}}
\caption{Scatter plot of the CP-odd TP asymmetry $A_{\mbox{\scriptsize
      CP}}^{\mbox{\scriptsize TP}} $ for various combinations of the
  NP parameters.}
\label{fig:ACPodd}
\end{center}
\end{figure}

Finally, we define the TPA as
\begin{eqnarray}
  A_{\mbox{\scriptsize CP}}^{\mbox{\scriptsize TP}} 
  \equiv \frac{\Gamma_{\mbox{\scriptsize TP}}
    -\overline{\Gamma}_{\mbox{\scriptsize TP}}}
   {\Gamma+\overline{\Gamma}},
\end{eqnarray}
so that
\begin{eqnarray}
  A_{\mbox{\scriptsize CP}}^{\mbox{\scriptsize TP}} 
   & \simeq & \frac{1}{\cal R}
   \frac{48\sqrt{2}G_F m_t^2}{35 \pi}
   \frac{
    \mbox{Im}\!\left[X_T\left(X^{S*}_{LL}+X^{S*}_{RR}\right)
       -2X_{TE}\left(X^{S*}_{LL}-X^{S*}_{RR}\right)\right]}
        {\left(1-\zeta_W^2\right)^2\left(1+2\zeta_W^2\right)} \nonumber \\
   & \simeq & \frac{0.243}{\cal R}\,
    \mbox{Im}\!\left[X_T\left(X^{S*}_{LL}+X^{S*}_{RR}\right)
       -2X_{TE}\left(X^{S*}_{LL}-X^{S*}_{RR}\right)\right] .
   \label{eq:ATPCPnon-suppressed}
\end{eqnarray}
Table~\ref{tab:ACPTPNPNP} contains some numerical results following
from the above expression, showing that the TPAs can indeed be
large -- of order 10's of percent -- if the NP coefficients
are assumed not to be suppressed.  Figure~\ref{fig:ACPodd} shows a
scatter plot of $A_{\mbox{\scriptsize CP}}^{\mbox{\scriptsize TP}}$
versus ${\cal R}$.  The points in this plot were obtained by
generating combinations of purely real and purely imaginary random
numbers for $X^S_{LL}$, $X^S_{RR}$, $X_T$ and $X_{TE}$ over various
ranges.  Once again, asymmetries were only plotted if ${\cal R}\leq
3$.  It is evident from the plot that relatively large TPAs are
possible.

\section{Discussion and Conclusions}
\label{sec:discussion}

In this paper we consider new-physics (NP) contributions to the decay
$t\to b\overline{b}c$. In the Standard Model (SM), this is a
tree-level process: $t\to b W\to b \overline{b}c$. However, the SM
amplitude involves the small Cabibbo-Kobayashi-Maskawa element
$V_{cb}$ ($\simeq 0.04$), and is therefore suppressed. As a result,
the decay is quite sensitive to NP. Rather than working within the
context of any one particular extension of the SM, we parameterize the
NP couplings by an effective Lagrangian that includes the 10 possible
four-Fermi operators.  We show that the SM and NP contributions to
$t\to b\overline{b}c$ can indeed be about the same size.

We first compute the $t \to b\overline{b}c$ partial width in the
presence of NP. The ratio ${\cal R}$, defined in Eq.~(\ref{eq:Rcal}),
provides a quantitative measure of the deviation of the partial width
from its SM expectation (${\cal R} = 1$ in the SM). This shows clearly
that this observable is excellent for showing that NP is present --
significant deviations of ${\cal R}$ from 1 are possible.

On the other hand, the partial width is not a good observable to {\it
  identify} the new physics -- all 10 NP operators contribute to
${\cal R}$ in a similar way. In order to get an idea of the type of NP
present, it is necessary to consider other quantities. To this end, we
construct two CP-conserving and two CP-violating observables: (i)
CP-even: a forward-backward-like asymmetry [$A_{\overline{\rho}^2}$ --
  Eq.~(\ref{eq:Arhobarsq})] and a top-quark-spin-dependent asymmetry
[$A_\parallel$ -- Eq.~(\ref{eq:Aparallel})], (ii) CP-odd: the partial
rate asymmetry [$A_{\mbox{\scriptsize CP}}$ -- Eq.~(\ref{eq:PRA1})]
and a triple-product asymmetry [$A_{\mbox{\scriptsize
      CP}}^{\mbox{\scriptsize TP}}$ --
  Eq.~(\ref{eq:ATPCPnon-suppressed})]. In each case, the observable is
formulated in such a way that it is zero within the context of the SM.
The key point is that these observables depend on differing combinations of the NP
parameters. This gives them different sensitivities to the various
Lorentz structures present in the NP effective Lagrangian.

The allowed values of these four observables vary greatly depending on
the values of the NP parameters, but results in the 10-20\% range are
possible (see Tables \ref{tab:ACPeven} and \ref{tab:ACPTPNPNP}). If NP
is present, it may well produce measurable values of these
observables. Taken together, the measurements of these quantities will
give a good indication of the type of NP present.

\bigskip
\noindent
{\bf Acknowledgments}:
%\bigskip 
We would like to thank A. Soni for helpful discussions.  This work was
financially supported by NSERC of Canada (D.L.)  The work of K.K.,
T.K. and K.W. was supported by the U.S.\ National Science Foundation
under Grant PHY--0900914.  
This work was also partially supported by ANPCyT (Argentina) 
under grant \# PICT-PRH 2009-0054.

%%%%%%%%%%%%%%%%%%%%%%%%%%%%%%%%%%%%%%%%%%%%%%%%%%%%%

\appendix

\section{\boldmath Useful Expressions for 
$t\to b\overline{b} c$}
\label{sec:AppA}

This Appendix contains two expressions that are used to compute
observables in the main body of the paper. We take $m_b = m_c \simeq
0$.  The first of these is the expression for the partial differential
decay width for $t\to \overline{b}bc$. Using Eqs.~(\ref{eq:Wamp}) and
(\ref{eq:NPamplitude2}) and averaging over the top quark's spins and colours, we
find the following,
\begin{eqnarray}
  &&\frac{d\Gamma}{dq^2d\rho^2}= \frac{3 G_F^2\left(V_{tb}V_{cb}\right)^2}
      {2\left(2\pi\right)^3m_t^3}
   \Bigg\{\left(q^2+\rho^2\right)\left(m_t^2-q^2-\rho^2\right) \nonumber\\
 && \times\Bigg[ m_W^4 \left|G_T\right|^2
    +4m_W^2\,\mbox{Re}\left(G_T X_{LL}^{V*}\right)
    +4\left(\left|X_{LL}^{V}\right|^2+\left|X_{RR}^{V}\right|^2\right)
    \Bigg]~~~\nonumber\\
 && +~4\rho^2\left(m_t^2-\rho^2\right)
   \left(\left|X_{LR}^{V}\right|^2+\left|X_{RL}^{V}\right|^2\right) 
   \nonumber\\
 && +~q^2\left(m_t^2-q^2\right)
   \left(\left|X_{LL}^{S}\right|^2+\left|X_{RR}^{S}\right|^2
      +\left|X_{LR}^{S}\right|^2+\left|X_{RL}^{S}\right|^2\right)
   \nonumber\\
 &&+~8q^2\left(-m_t^2+q^2+2\rho^2\right)
   \mbox{Re}\left[X_T\left(X_{LL}^{S*}+X_{RR}^{S*}\right)
     -2X_{TE}\left(X_{LL}^{S*}-X_{RR}^{S*}\right)\right] \nonumber\\
 &&+~32 \left[m_t^2\left(q^2+4\rho^2\right)-\left(q^2+2\rho^2\right)^2\right]
   \left(\left|X_T\right|^2+4\left|X_{TE}\right|^2\right)
   \Bigg\} ,
\label{eq:AppdGam}
\end{eqnarray}
in which $V_{tb}$ and $V_{cb}$ have been taken to be real.  The
analogous expression for $\overline{t}\to\overline{b}b\overline{c}$ is
obtained by complex conjugating all of the NP coefficients
($X_{LL}^V$, etc.), while leaving $G_T$ unchanged.

It is also useful to have the expression for the absolute value
squared of the amplitude, keeping only the terms that contain the spin
four vector for the top quark.  This expression is used to compute the
CP-even single-spin asymmetry in Sec.~\ref{sec:SingleSpin} and the TP
asymmetry in Sec.~\ref{sec:TPAsym}.  Keeping only terms containing the
spin four vector of the top quark, we find,
\begin{eqnarray}
 &&\frac{1}{3}  
   \sum_{\mbox{\scriptsize colours}}
   \sum_{~b, \overline{b}, c\mbox{\scriptsize ~spins}}
   \left.\left| {\cal M}\right|^2\right|_{s_t} \nonumber\\
     &&= 96 G_F^2m_t\left(V_{tb}V_{cb}\right)^2 
\Bigg\{\left(m_t^2-q^2-\rho^2\right) 
  \Bigg[ -m_W^4 \left|G_T\right|^2
    -4m_W^2\,\mbox{Re}\left(G_T X_{LL}^{V*}\right) \nonumber\\
 &&
    -~4\left(\left|X_{LL}^{V}\right|^2-\left|X_{RR}^{V}\right|^2\right)
    \Bigg] p_{\overline{b}}\cdot s_t
  -4\rho^2\left(\left|X_{LR}^{V}\right|^2-
       \left|X_{RL}^{V}\right|^2\right) p_c\cdot s_t \nonumber\\
 && 
       -~q^2 \left(\left|X_{LL}^{S}\right|^2-\left|X_{RR}^{S}\right|^2
      +\left|X_{LR}^{S}\right|^2-\left|X_{RL}^{S}\right|^2\right)
      p_{b}\cdot s_t
   \nonumber\\
 &&+~8\,\mbox{Re}\left[X_T\left(X^{S*}_{LL}-X^{S*}_{RR}\right)
      -2X_{TE}\left(X^{S*}_{LL}+X^{S*}_{RR}\right)\right]
   \left[\left(m_t^2-q^2\right)p_{\overline{b}}\cdot s_t
     +\rho^2 p_b\cdot s_t \right]
    ~~~\nonumber\\
 &&+~128 \,\mbox{Re}\left[X_T X_{TE}^*\right]
   \left[\left(2m_t^2 -q^2-2\rho^2\right)p_{\overline{b}}\cdot s_t
     +\left(q^2+2\rho^2\right) p_c\cdot s_t\right]\nonumber \\
 &&-~16\,\mbox{Im}\left[X_T\left(X^{S*}_{LL}+X^{S*}_{RR}\right)
      -2X_{TE}\left(X^{S*}_{LL}-X^{S*}_{RR}\right)\right]
   \epsilon\left(p_t, s_t, p_{\overline{b}},p_c\right)
    \Bigg\},
  \label{eq:absMsq_top_spin_only}
\end{eqnarray}
in which $s_t$ denotes the top's spin four vector and
$\epsilon\left(p_t, s_t, p_{\overline{b}},p_c\right)\equiv
\epsilon_{\alpha\beta\gamma\delta} p_t^\alpha s^\beta_t
p_{\overline{b}}^\gamma p_c^\delta$.  In writing the above expression,
we have used the fact that $p_t\cdot s_t=0$.  We have also summed over
quark colours and over the final-state quarks' spins, and have divided
by 3 for the average over the top quark's colours.

\section{CPT and Beyond}
\label{sec:CPT}

The CPT theorem requires the total decay width for the top to be equal
to that for the anti-top.  An apparent violation of the CPT theorem
arises, however, if the NP contributions in
Fig.~\ref{fig:feynman_diagram1} (b) are the only ones that are kept.
That this is the case is straightforward to see, since the diagram in
Fig.~\ref{fig:feynman_diagram1} (b) affects the partial widths for
$t\to b\overline{b} c$ and $\overline{t} \to \overline{b} b
\overline{c}$ differently [leading to the PRA in Eq.~(\ref{eq:PRA1})],
but has no effect on the other top or anti-top decay modes.  Thus, the
top and anti-top total widths are not equal if only such contributions
are kept, resulting in an apparent violation of the CPT theorem.  This
phenomenon is well-known (see, for example, Refs.~\cite{gerard1988,
  wolfenstein1990, Eilam, soares,Eilamreply,arens}).  In this Appendix
we show that the inclusion of certain vertex-type corrections gives
rise to compensating differences in the top and anti-top widths.  The
sum of the differences is zero, so that the top and anti-top widths
no longer differ, in agreement with the CPT theorem.

Let us define the partial width difference for the decay $t\to b
\overline{j} k$ as follows,
\begin{eqnarray}
  \Delta \Gamma\left(t\to b \overline{j} k\right) \equiv 
    \Gamma\left(t\to b \overline{j} k\right) -
    \Gamma\left(\overline{t}\to \overline{b}j \overline{k}\right),
\end{eqnarray}
in which $j$ and $k$ could refer either to quarks or to leptons.  For
the case $t\to b\overline{b}c$, the main contribution to
$\Delta\Gamma$ is due to the interference between the SM and NP
diagrams indicated in Fig.~\ref{fig:feynman_diagram1}.  Another
important set of contributions for the decay $t\to b \overline{j}k$ is
indicated in Fig.~\ref{fig:CPT_vertex_correction}.  The absorptive
parts of these vertex-like corrections interfere with their associated
SM diagrams in such a way that the conservation of CPT is manifest.
Using the Cutkosky rules to calculate the absorptive part of the
vertex-like corrections, we find
\begin{eqnarray}
  \Delta\Gamma\left(t\to b \overline{j}k\right)
    \simeq \frac{2\sqrt{2}G_F m_W^2\left(V_{cb}\right)^2}{\pi}
    \mbox{Im}\!\left(X^{V*}_{LL}\right)\!
    \Gamma\left(t\to b W\right)\!\left[\delta_{\overline{j}\,\overline{b}}
      \delta_{kc} 
      \!-\!{\cal B}\left(W\to \overline{j}k\right)\right]\!.~~~
    \label{eq:DelGamma}
\end{eqnarray}
Summing over $\overline{j}$ and $k$ (including both quark and lepton
final states), we have
\begin{eqnarray}
  \sum_{\overline{j},k} \Delta\Gamma\left(t\to b \overline{j}k\right)
  = 0 ,
\end{eqnarray}
demonstrating that the CPT theorem is indeed respected once the
absorptive parts of the diagrams in
Fig.~\ref{fig:CPT_vertex_correction} are included.

Equation~(\ref{eq:DelGamma}) gives a correction to the PRA for $t\to b
\overline{b}c$, leading to the following modification of
Eq.~(\ref{eq:PRA1}),
\begin{eqnarray}
  A_{\mbox{\scriptsize CP}}
  \simeq 0.102\times
   \frac{\mbox{Im}\left(X_{LL}^{V*}\right)}{\cal R} \left[1-
     {\cal B}\left(W\to \overline{b}c\right)\right] .
   \label{eq:PRA2}
\end{eqnarray}
The correction to the original expression is miniscule, since ${\cal
  B}\left(W\to \overline{b}c\right)\simeq \left|V_{cb}\right|^2/3
\simeq 5.5\times 10 ^{-4}$.

An interesting consequence of the CPT theorem is that, if NP operators
give a PRA in a particular decay mode (such as $t\to b \overline{b}
c$, as in our case), then those same NP operators must also contribute
to one or more other decay modes in such a way that the total width of
the top is the same as that of the anti-top.  This means that those
other decay modes must also have PRAs (barring other accidental
cancellations).  We can use Eq.~(\ref{eq:DelGamma}) to estimate the
PRAs in other decay modes due to the NP operators in
Eqs.~(\ref{eq:eff1})-(\ref{eq:eff3}).  The resulting expression is
given by,
\begin{eqnarray}
  A_{\mbox{\scriptsize CP}}\!\left(t\to b \overline{j}k\right)
  \!\simeq\! -\frac{\sqrt{2}G_Fm_W^2\!\left(V_{cb}\right)^2}{\pi}
  \mbox{Im}\!\left(X_{LL}^{V*}\right)
  \!\simeq\! -5.6\times 10^{-5}\,\mbox{Im}\!\left(X_{LL}^{V*}\right) ,
  ~\overline{j}k\neq \overline{b}c .~~
   \label{eq:PRA3}
\end{eqnarray}
Thus the contributions of these NP operators to PRAs in other top
decay modes are expected to be very small.  One could also consider
the complementary question: Are there NP operators, other than those
given in Eqs.~(\ref{eq:eff1})-(\ref{eq:eff3}), that could contribute
to the PRA in $t\to b\overline{b}c$?  The answer to this question
appears to be yes.  For example, the effective operators
$\left(\overline{s}\,{\cal O}_1c\right)\!\left(\overline{c}\,{\cal
  O}_2 b\right)$ or $\left(\overline{d}\,{\cal
  O}_1u\right)\!\left(\overline{c}\,{\cal O}_2 b\right)$ could appear
in a diagram similar to that in Fig.~\ref{fig:CPT_vertex_correction},
but with the usual SM $tbW$ vertex at the top, and the NP-induced
one-loop correction to the $Wbc$ vertex at the bottom.  Such operators
are constrained by $B$ decays.

\begin{figure}[t]
\begin{center}
\resizebox{3.5in}{!}{\includegraphics*{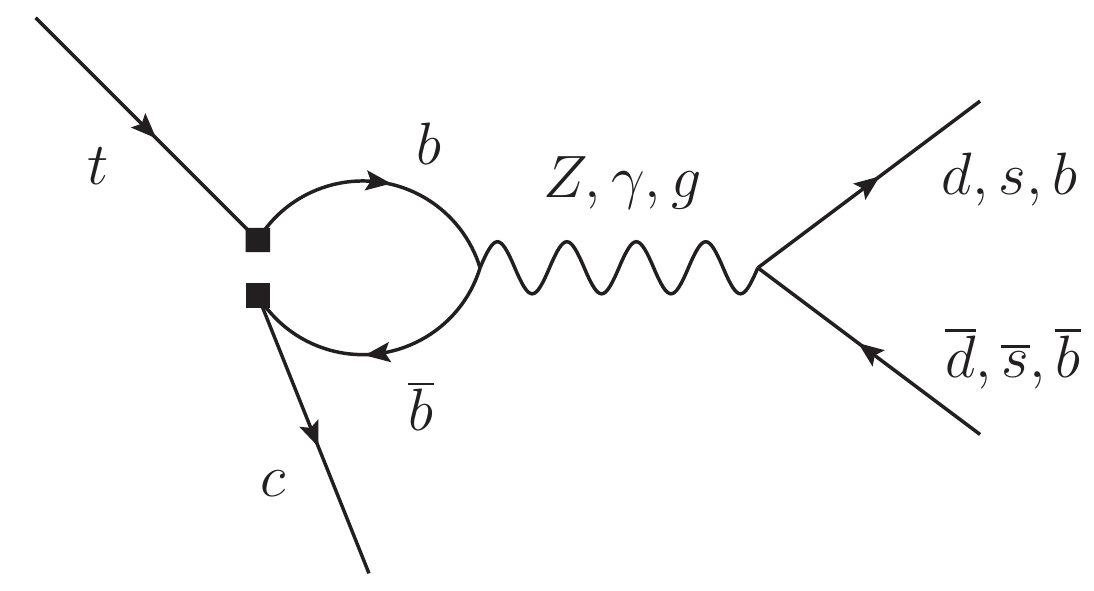}}
\caption{A loop-level contribution of the NP operators that could
contribute to PRAs in $t\to f\overline{f}c$, with $f=d, s, b$.}
\label{fig:tcZvertex}
\end{center}
\end{figure}

We should note that the NP effective operators in
Eqs.~(\ref{eq:eff1})-(\ref{eq:eff3}) give rise to other loop-level
diagrams that could contribute to PRAs in top decays.  The
contributions in different decay modes would still complement each
other in the sense that the total top and anti-top widths would remain
equal.  Figure~\ref{fig:tcZvertex} shows an example of loop-level
corrections to $t\to f\overline{f}c$ ($f=d,s,b$) mediated by the NP
operators considered in this work.  These diagrams could interfere
with their corresponding SM diagrams to induce PRAs.  We do not
compute such contributions here.

\section{Effect of Including Colour-mismatched Terms}
\label{sec:AppB}

The effective Lagrangian incorporating NP effects given in
Eqs.~(\ref{eq:eff1})-(\ref{eq:eff3}) assumed that the colour indices
contracted in the same manner as those of the SM diagram.  This need
not be the case, so it is useful to consider the effects of including
colour-mismatched terms in the effective Lagrangian.  To this end,
let us generalize the NP effective Lagrangian in
Eq.~(\ref{eq:eff1}) as follows,
\begin{eqnarray}
  {\cal L}_{\mbox{\scriptsize eff}}^V & = & \frac{g^{\prime 2}}{M^2}\left\{
    {\cal R}_{LL}^V\,\overline{b}_a\gamma_\mu P_L t_a 
      \,\overline{c}_b\gamma^\mu P_L b_b
   + {\cal R}_{LL}^{V\prime}\,\overline{b}_a\gamma_\mu P_L t_b 
      \,\overline{c}_b\gamma^\mu P_L b_a
\right.\nonumber\\
& &~~~~\left.
   + {\cal R}_{LR}^V\,\overline{b}_a\gamma_\mu P_L t_a 
      \,\overline{c}_b\gamma^\mu P_R b_b
   + {\cal R}_{LR}^{V\prime}\,\overline{b}_a\gamma_\mu P_L t_b 
      \,\overline{c}_b\gamma^\mu P_R b_a
\right.\nonumber\\
& &~~~~\left.
   + \ldots\right\} + \mbox{h.c.,}
\end{eqnarray}
and similarly for Eqs.~(\ref{eq:eff2}) and (\ref{eq:eff3}).  In this
expression, the subscripts $a$ and $b$ are colour indices and the
primed coefficients correspond to the new, colour-mismatched terms.
The total amplitude for $t_a\to b_b\overline{b}_c c_d$ (with the
subscripts $a$, $b$, $c$ and $d$ representing the colours) could then
be parameterized as
\begin{eqnarray}
  {\cal M}_{abcd} = \sum_i\left(\mathbb{R}_i\delta_{ab}\delta_{cd}
   +\mathbb{R}_i^\prime\delta_{ad}\delta_{bc}\right){\cal M}_i\, ,
   \label{eq:Mabcd}
\end{eqnarray}
in which the sum runs over the SM diagram, plus all NP contributions.
The factors $\mathbb{R}_i$ and $\mathbb{R}_i^\prime$ are the
coefficients for the colour-matched and colour-mismatched terms,
respectively, and are assumed to contain all of the weak phases.  (The
$\mathbb{R}^\prime$ coefficient for the SM term is assumed to be
zero.)  For a given value of $i$, the phases of $\mathbb{R}_i$ and
$\mathbb{R}_i^\prime$ could be different.  The factors ${\cal M}_i$
contain all the spinors and $\gamma$ matrices and, in the case of the
SM diagram, the $W$ propagator.

Summing over the quarks' colours and dividing by 3 for the average
over the top quark's colours, we find,
\begin{eqnarray}
  \frac{1}{3}\!\sum_{a,b,c,d}\!{\cal M}_{abcd}{\cal M}_{abcd}^*
  & = & 3 \sum_i\left[
    \left|\mathbb{R}_i\right|^2
    + \left|\mathbb{R}_i^\prime\right|^2
    + \frac{2}{3}\,\mbox{Re}\!\left(\mathbb{R}_i\mathbb{R}_i^{\prime *}
    \right) \right]\left|{\cal M}_i\right|^2 \nonumber\\
  &+& 6 \sum_{j>i} 
   \left\{\mbox{Re}\!\left[\mathbb{R}_i\mathbb{R}_j^* +
   \mathbb{R}_i^\prime \mathbb{R}_j^{\prime *} +
   \frac{1}{3}\left(
   \mathbb{R}_i\mathbb{R}_j^{\prime *} +
   \mathbb{R}_i^\prime\mathbb{R}_j^*\right)\right]
   \!\mbox{Re}\!\left({\cal M}_i {\cal M}_j^*\right) \right. \nonumber\\
  &-& \left. \mbox{Im}\!\left[\mathbb{R}_i\mathbb{R}_j^* +
   \mathbb{R}_i^\prime \mathbb{R}_j^{\prime *} +
   \frac{1}{3}\left(
   \mathbb{R}_i\mathbb{R}_j^{\prime *} +
   \mathbb{R}_i^\prime\mathbb{R}_j^*\right)\right]
   \!\mbox{Im}\!\left({\cal M}_i {\cal M}_j^*\right) \right\} .
   \label{eq:colourmismatchcorrection}
\end{eqnarray}
The $\mathbb{R}_i\mathbb{R}_j^*$ terms in the above expression
correspond to the ``colour-matched'' terms that we have taken into
account in this work.  The other terms are new.

Equation~(\ref{eq:colourmismatchcorrection}) can be used to generalize
the expressions in this paper, provided the expressions have already
been split cleanly into pieces containing the weak phases
($\mathbb{R}_i$, etc.) and those containing the spinors and any strong
phases (${\cal M}_i$).  Expressions containing SM-NP cross-terms may
safely set $\mathbb{R}_{\mbox{\scriptsize SM}}=1$ and incorporate the
entire amplitude into the ``${\cal M}_{\mbox{\scriptsize SM}}$'' part
[in Eq.~(\ref{eq:Mabcd})], since $V_{tb}$ and $V_{cb}$ have been taken
to be real.  As an example, the generalized form for
Eq.~(\ref{eq:Rcal}) would be
\begin{eqnarray}
  {\cal R} \simeq 1 + 0.0845\times \Bigg[&&\!\!\!\!\!
    -0.05\!\times\!\mbox{Re}\left(X_{LL}^{V*}
    +\frac{1}{3}X_{LL}^{V\prime*}\right) \nonumber\\
    & & +
    \left|X^V_{LL}\right|^2+
    \left|X^{V\prime}_{LL}\right|^2+ 
    \frac{2}{3}\mbox{Re}\left(X^{V}_{LL}X^{V\prime *}_{LL}\right)
    + \ldots \Bigg]\, ,
\end{eqnarray}
in which we have used the fact that $\mathbb{R}_{\mbox{\scriptsize
    SM}}^\prime=0$.  Similarly, Eq.~(\ref{eq:ATPCPnon-suppressed})
would become,
\begin{eqnarray}
  A_{\mbox{\scriptsize CP}}^{\mbox{\scriptsize TP}} 
  & \simeq & \frac{0.243}{\cal R}\,
    \mbox{Im}\!\left\{X_T\left(X^{S*}_{LL}+X^{S*}_{RR}\right)
       -2X_{TE}\left(X^{S*}_{LL}-X^{S*}_{RR}\right) \right.\nonumber \\
  & &~~~~~~~+ \left.
       X_T^\prime\left(X^{S\prime *}_{LL}+X^{S\prime *}_{RR}\right)
       -2X_{TE}^\prime\left(X^{S\prime *}_{LL}-X^{S\prime *}_{RR}\right) 
       \right.\nonumber \\
  & &~~~~~~~ \left.
       + \frac{1}{3}\left[
       X_T\left(X^{S\prime *}_{LL}+X^{S\prime *}_{RR}\right)
       -2X_{TE}\left(X^{S\prime *}_{LL}-X^{S\prime *}_{RR}\right) 
       \right.\right. \nonumber \\
  & &~~~~~~~ \left.\left.
      + X_T^\prime\left(X^{S *}_{LL}+X^{S *}_{RR}\right)
       -2X_{TE}^\prime\left(X^{S *}_{LL}-X^{S *}_{RR}\right) 
       \right]\right\},
\end{eqnarray}
and Eq.~(\ref{eq:PRA1}) would become,
\begin{eqnarray}
  A_{\mbox{\scriptsize CP}} \simeq \frac{0.102}{\cal R}\times
   \mbox{Im}\left(X_{LL}^{V*}+\frac{1}{3}
     X_{LL}^{V\prime *}\right) .
\end{eqnarray}
Finally, an expression such as ``Re$\left(G_T
X_{LL}^{V*}\right)$'' in Eq.~(\ref{eq:AppdGam}), which contains both a
strong phase (in $G_T$) and a weak phase (in $X_{LL}^V$), first needs
to be separated into two pieces using
$\mbox{Re}\left(AB\right)=\mbox{Re}(A)\mbox{Re}(B)-
\mbox{Im}(A)\mbox{Im}(B)$.

%%%%%%%%%%%%%%%%%%%%% REFERENCES %%%%%%%%%%%%%%%%%%%%%%%

\end{document}